%% file: 0_Main_and_Abstract.tex
\documentclass{article}
\usepackage{tablefootnote}
\usepackage[lofdepth,lotdepth]{subfig}
\usepackage{arxiv}
\usepackage{graphicx}
\usepackage[utf8]{inputenc} % allow utf-8 input
\usepackage[T1]{fontenc}    % use 8-bit T1 fonts
\usepackage[hide links]{hyperref}       % hyperurl's
\usepackage{url}            % simple URL typesetting
\usepackage{booktabs}       % professional-quality tables
\usepackage{amsfonts}       % blackboard math symbols
\usepackage{amsmath}        
\usepackage{nicefrac}       % compact symbols for 1/2, etc.
\usepackage{microtype}      % microtypography
\usepackage{lipsum}
\usepackage[symbol]{footmisc}
\usepackage{pdfpages}
\usepackage{bm}
\usepackage{titlesec}
\titleformat{\paragraph}
{\normalfont\normalsize\bfseries}{\theparagraph}{1em}{}
\titlespacing*{\paragraph}
{0pt}{3.25ex plus 1ex minus .2ex}{1.5ex plus .2ex}
\usepackage{amsmath}
\usepackage{algorithm}
\usepackage[noend]{algpseudocode}
\usepackage{cite} 
\makeatletter
\g@addto@macro{\UrlBreaks}{%
\do\-}
\makeatother
\usepackage{arxiv}
\usepackage[utf8]{inputenc} % allow utf-8 input
\usepackage[T1]{fontenc}    % use 8-bit T1 fonts
\usepackage{hyperref}       % hyperlinks
\usepackage{url}            % simple URL typesetting
\usepackage{booktabs}       % professional-quality tables
\usepackage{amsfonts}       % blackboard math symbols
\usepackage{nicefrac}       % compact symbols for 1/2, etc.
\usepackage{microtype}      % microtypography
\usepackage{lipsum}
\usepackage{graphicx}
\graphicspath{ {images/} }
\usepackage{enumitem}

\usepackage{float}
\usepackage{mathtools}

\usepackage{tikz}
\usepackage{tikz-dependency}
\usetikzlibrary{fit,positioning,arrows.meta,shapes}
\tikzset{neuron/.style={shape=circle, minimum size=1.25cm, 
  inner sep=0, draw, font=\small}, io/.style={neuron, fill=gray!20}}
\usetikzlibrary{shapes,arrows}
\usetikzlibrary{calc}
\usetikzlibrary{positioning}

{
\title{Artificial Intelligence for Trading Strategies}
  \author{ Danijel Jevtic \\
  School of Engineering\\
  Zurich University of Applied Sciences\\
  Winterthur, Switzerland \\
  \texttt{jevtidan@students.zhaw.ch } \\
  \And
  Romain Délèze \\
  School of Engineering\\
  Zurich University of Applied Sciences\\
  Winterthur, Switzerland \\
  \texttt{delezrom@students.zhaw.ch} \\
  \And
  Joerg Osterrieder\footnotemark[1] \\
  School of Engineering\\
  Zurich University of Applied Sciences\\
  Winterthur, Switzerland \\
  \texttt{joerg.osterrieder@zhaw.ch} \\
}
}

\begin{document}
\maketitle
\footnotetext[1]{\tiny{Financial support by the Swiss National Science Foundation within the project “Mathematics and Fintech - the next revolution in the digital transformation of the Finance industry” is gratefully acknowledged by the corresponding author. 
This research has also received funding from the European Union's Horizon 2020 research and innovation program FIN-TECH: A Financial supervision and Technology compliance training programme under the grant agreement No 825215 (Topic: ICT-35-2018, Type of action: CSA).
Furthermore, this article is based upon work from the COST Action 19130 Fintech and Artificial Intelligence in Finance, supported by COST (European Cooperation in Science and Technology), www.cost.eu (Action Chair: Joerg Osterrieder).\newline
The authors are grateful to Stephan Sturm, Moritz Pfenninger, Samuel Rikli, Bigler Daniel, Antonio Rosolia, management committee members of the COST (Cooperation in Science and Technology) Action Fintech and Artificial Intelligence in Finance as well as speakers and participants of the $5^{\text{th}}$ and $6^{\text{th}}$ European COST Conference on Artificial Intelligence in Finance and Industry, which took place at Zurich University of Applied Sciences, Switzerland, in September 2020 and 2021.}}

\newpage
%\includepdf[pages=-]{SELBS_ERKL_BA.pdf}
\newpage
\begin{abstract}

In recent years, much research has been done in predicting stock price trends using machine learning algorithms. However, machine learning methods are often criticized by financial practitioners. They argue that neural networks are a "black box." In this bachelor thesis, we show how four different machine learning methods (Long Short-Term Memory, Random Forest, Support Vector Machine Regression, and k-Nearest Neighbor) perform compared to already successfully applied trading strategies such as Cross Signal Trading and a conventional statistical time series model ARMA-GARCH. The aim is to show that machine learning methods perform better than conventional methods in the crude oil market when used correctly. A more detailed performance analysis was made, showing the performance of the different models in different market phases so that the robustness of individual models in high and low volatility phases could be examined more closely. The focus was on the robustness and performance of the machine learning methods. It was essential to keep complexity in proportion to performance. All analysis work was realized with the software Python. It was shown that the machine learning methods such as Random Forest, on average, made better decisions in highly volatile market periods than ordinary statistical models such as ARMA-GARCH. The Random Forest was characterized by a very high true-positive rate (recall), which set it apart from the other models. Likewise, the final performance result provided conclusions about the robustness of the individual models. It could be shown in which market phases the models made good or bad trading decisions. In the case of the k-Nearest Neighbor model, which failed to detect the Corona crisis, it was shown that not every machine learning model responded well to different market phases. Comparing this with the Random Forest model, it can be seen that the most significant monthly gain was achieved there across all models.
The conventional statistical models - ARMA-GARCH - performed surprisingly well and were on par with the machine learning models. Nevertheless, the extreme values were analyzed in more detail. It was found that the Random Forest model made the best decisions at the 95\% interval level. This shows that the Random Forest model always made better decisions than all other models during periods when a lot is going on in markets, which is reflected in the total return. Surprisingly, the ARMA-GARCH model made the best decisions at the 99\% level but performed worse on average when the market was "calmer." It was shown that better results could be achieved with proper machine learning methods than with conventional statistical methods. For further investigation, these models would also have to be analyzed in other markets.


\end{abstract}
\maketitle
\newpage
\tableofcontents
\newpage

% Abstract is in this File
\include{1_Introduction}
\include{2_Literature_Overview}

\include{3_Data}

\include{4_Experiment}
\include{5_Conclusion}

\include{6_References}
\bibliographystyle{IEEEtran}
\bibliography{references} 
\end{document}

%% file: 1_Introduction.tex
\section{Introduction} \label{Section: Introduction}

Our research combines three main building blocks: Financial Time Series, Machine Learning, and Trading Strategies. We briefly reviewed existing research in each of these three areas. We used a classical approach with an ARMA-GARCH model and Cross Signal Trading as a benchmark in this work. We then compared the performance of different machine learning models such as Random Forest (RF), k-nearest neighbors (kNN), Support Vector Machines (SVM), and a long short-term memory (LSTM) Network

\subsection{Financial Markets and Data}

Less than 70 years after the term was coined, artificial intelligence has become an integral part of the most sophisticated and fast-moving industries, with significant impacts for the financial industry. The impact of AI and ML tools on financial markets cannot be underestimated. Strong computational power allows large amounts of data to be processed in a short period of time, while cognitive computing helps manage structured and unstructured data. Algorithms analyze financial data and identify early signs of potential future problems. Financial markets, like the economy, are highly complex systems, and it is often impossible to explain macroscopic phenomena by simple summaries of microscopic processes or events. The outcomes of actions in the system are difficult to predict, and sometimes it is impossible to find the cause of significant anomalies or even the factors that influence the event. \cite{cdi_proquest_reports} AI in finance is a powerful ally in analyzing real-time activity in a given market or environment, providing accurate and detailed forecasts based on multiple variables that are critical to business planning. For example, we mentioned Equbot's AI-driven equity fund ETF (AIEQ), which is based on IBM's Watson AI program. AIEQ is the first ETF to use artificial intelligence for stock selection. AI autonomously analyzes data, processes it, and then makes investment decisions. AI can beat the market without the deployment delays of traditional data science methods. \cite{baron_2021} 

Financial data consists of pieces or sets of information related to the financial health of a business. The pieces of data are used by internal management to analyze business performance and determine whether tactics and strategies must be altered. People and organizations outside a business will also use financial data reported by the business to judge its credentials, decide whether to invest in the business and determine whether it complies with government regulations. \cite{grimsley_2021}

\subsection{Financial time-series modelling}

The financial industry has always been interested in successfully predicting financial time series. Numerous studies based on conventional time series analysis and ML models have been published. Due to the successful research in AI, more and more conventional models are combined with ML models, which have been met with success.

There are several methods of modeling financial time series. In particular, the ARCH/GARCH models, which are based on classical statistics, model the change in variance over time in a time series by describing the variance of the current error term as a function of past errors. Often, the variance is related to the squares of the previous innovations as AR (Auto-Regressive) models rely heavily on past information to make a prediction. They are often used in time series models where time-varying volatility and volatility clusters are present. We will primarily rely on the ARCH/GARCH models as the statistics are trackable and, therefore, can be considered a good benchmark. \cite{francq_zakoian_2019}

\subsection{Neural networks and Machine learning}

Neural networks are a subset of machine learning techniques and are at the heart of deep learning algorithms. Deep neural networks consist of node layers that contain three different layers that are input layers, hidden layers, and output layers. In the simplest case, each node is connected to another and has an associated weight and threshold. If the output of a single perceptron is above the specified threshold value, that perceptron is activated and sends data to the next layer of the network. Otherwise, no data is forwarded to the next layer of the network.

Neural networks rely on training data to learn and improve their accuracy over time. Once these learning algorithms achieve high accuracy, they are potent tools in computer science and artificial intelligence, allowing us to classify cluster data and serve as universal approximation tools for unknown functions. Machine learning models such as RF are typically treated as black boxes. Due to the fact that a forest consists of many deep trees, where each tree is trained on bagged data using a random selection of features, gaining a complete understanding by examining each tree would be close to impossible. \cite{wijaya_2021}

\subsection{Research Contribution}

This paper provides an overview of the current machine learning models used in finance. The aim is to explain the information gained from trading strategies using different ML models. The data of Brent crude Oil to different ML models will be analyzed and applied to trading strategies. Furthermore, the functionality of ML models will be taught, and the current state of research and practice will be presented in applications.

\subsection{Structure of this paper}

The continuation of the thesis is structured as follows: Section 2 will present the background literature on artificial intelligence for trading strategies, current research, and similar approaches while section 3 introduces the experimental work (a weak form of EMH). Lastly, section 5 will conclude the thesis and provide an outlook at further research and applications.

%% file: 2_Literature_Overview.tex
\section{Literature Overview} \label{Section: Literature Overview}

In the literature overview, the used literature is briefly summarized. In the beginning, the Efficient Markets and Rational Expectations are discussed. Afterward, machine learning in Finance is introduced. In the chapter Background, the used models are discussed to overview the upcoming work.

\subsection{Efficient markets and Rational Expectations}

The theory of efficient markets was established in 1970 by Eugene F. Fama (Efficient Capital Markets: A Review of Theory and Empirical Work). Fama described efficiency as "Markets are efficient only if prices always "fully reflect" all available information." Fama divided the efficient market theory into three different forms: the weak form, which includes historical prices, and the semi-strong form, which identifies with public information such as profits/losses, stock splits, etc. The strong form includes insider information. Further in itself, the question was whether the market prices would last exactly. However, this stands in conflict with whether a constant outperformance would be possible; this is not possible alone by using already known information. Other conditions were defined so that the capital market efficiency was fulfilled. The first condition: no transaction costs, the second: all information is available to all market participants at no cost, and the third: all agree on the impact of current information on the current price and distribution of future prices of each security. \cite{fama}

Charlie Munger quotes: "I think it is broadly true that the market is efficient, which makes it very hard to beat. But I do not think it is completely efficient." \cite{shiller_2003}

Therefore, a perfectly efficient market can be unpredictable because the more efficient it is, the more random and unpredictable the returns. Due to this, most economists agree that the EMH is not absolute, which Eugen Fama also agreed with. The participants, like the investors or simple traders, form the market, so everyone constantly influences the efficiency of the market. The public tries to gain a profit from information advantages. This is why the thesis is that prices contain all information reflecting the intrinsic value of the underlying asset. Additionally, the EMH does not work perfectly in the real world because people are irrational. Therefore, papers were also written regarding rational expectations. In 1961, John F. Muth said that "Our hypothesis is based on that dynamic economic models do not assume sufficient rationality." This remark should be understood that not perfect efficiency of the markets as a conclusion. \cite{sinclair_2008}

\subsection{Machine Learning in Finance}

The financial industry is most simply described as an information-processing industry, and each sector processes information differently. For example, investment funds make their investment decisions by processing information, while insurance companies use the information to determine the price of future insurance policies. In the 21st century, large amounts of data are generated daily. Therefore, it is obvious to use machine learning or neural networks for information processing. Now the question arises, what is machine learning? To answer this, we cite Arthur Samuel's quote published in 1959: "Machine learning is the subfield of computer science that gives computers the ability to learn without being explicitly programmed." In other words, machine learning does not require conventional rules, which humans have programmed in all software. It gives the computer the ability to learn on its own. We let the computer develop its own rules through pattern recognition. There are four different types of learning: supervised, unsupervised, reinforcement learning, and deep learning.

In supervised learning, functions are determined based on training data whose output is known (classification and regression). In unsupervised learning, patterns are recognized from unknown data sets, and rules are derived (clustering as one of the primary forms). In reinforcement learning, the algorithm is given parameters that it can control and vary, and it optimizes towards this goal independently through simulations. Finally, deep learning - is a system of artificial software neurons based on the human brain, which communicate with each other and learn independently from data. \cite{cfi_2022}

\newpage

\subsection{Background}

In this chapter, the used models are brought closer. The basic ideas of the process or algorithm are explained and further discussed in the chapter titled Experiment.  

\textbf{ARMA-GARCH Models} 

One of the most challenging and complex topics in finance is modeling financial time series. The problems can be attributed to the complexity and variety of stocks, currency exchange rates, interest rates, and the importance of observation frequencies such as seconds, minutes, hours, or days. Often the availability of large data sets is also a problem. However, the biggest problem is attributed to the existence of statistical regularities (stylized facts) that are common to many financial series and are difficult to reproduce artificially with stochastic models. Most of these stylized facts were presented in a scholarly article by Mandelbrot (1963).

In general, the objective of time series analysis is to find a model for the underlying stochastic process. This model is then used to analyze the process's structure or make optimal predictions. The class of ARMA models is most commonly used to predict steady-state processes. The problem is that the returns of financial time series depends on time and prove volatility clusters. For this problem, the ARCH model was developed by Engel in 1982, which can model autoregressive, conditionally heteroskedastic patterns in the data. A few years later, the generalized ARCH model GARCH was developed by Bollerslev (1986). In these models, the key concept is the conditional variance, i.e., the variance that depends on the past. The conditional variance is expressed as a linear function of the squared past values of the series. Now one can capture the vital stylized facts that characterize the financial series. \cite{francq_zakoian_2019_2}

\textbf{Random Forest}

RF is an algorithm used for classification and regression tasks. To have a more analyzed view of what RF is, we refer to the book written by James et al. (An Introduction to Statistical Learning), published in 2013. It combines the results of many different decision trees to make the best possible decisions. The learning algorithm belongs to the supervised learning methods. This uses the results of many different decision trees to make the best possible decisions or predictions. The decision trees were created randomly in an uncorrelated manner. Each tree makes individual decisions for itself. From the set of individual decisions, the algorithm delivers a final decision.

In order to fully understand the functional principle of RF, the terms decision tree and bagging must first be explained in more detail. Decision trees form the basis for RF and a single tree consists of several branches. The branches are then created by assigning data to a class based on their properties using rules.  Beginning from the first decision, more and more branches are created until a certain result level is reached. Bagging is a specific method of combining the individual predictions of different classification models. The individual results of the decision trees are included in the overall result with a specified weighting. \cite{liaw2002classification}

\textbf{Support Vector Machines}

SVM are generally used for classification and regression problems. Furthermore, SVM can easily handle multiple continuous and categorical variables. In order to separate the data set into different classes, SVM constructs a hyperplane in multidimensional space. To guarantee the minimization of the error, SVM iteratively generates an optimal hyperplane. The core idea is to find a maximum marginal hyperplane that best classifies the data-set. The support vectors, hyperplane, and the margin have to be explained to understand the procedure better.

Support vectors are the data points that are closest to the hyperplane. These points have the most significant impact on the classification of the data. Through these points, the dividing line is better defined by calculating the margins. The hyperplane can be seen as a decision plane. This is between the set of objects lies, with which the different class memberships are separated. The margin is the gap between the two lines at the closest class points. The margin is the perpendicular distance between the line and the support vectors. The bigger the gap between the classes, the better. A smaller gap is considered flawed. \cite{smola_schoelkopf_2004}

\newpage

\textbf{K-nearest-neighbors}

The kNN algorithm belongs to the supervised machine learning methods and is one of the topmost machine learning algorithms. kNN is used in various applications such as finance, healthcare, political science, and credit ratings. It can be used to solve classification and regression problems. Furthermore, the kNN algorithm is based on the feature similarity approach.

The focus of this section is to explain how the algorithm works and how it is defined. In supervised learning, the algorithm is given a data-set labeled with appropriate output values with which it can train and define predictive models. Subsequently, the trained model will be applied to a test set to predict the corresponding output values. The intuition of the kNN algorithm is simple to understand. kNN is a non-parametric learning algorithm. Non-parametric means that there is no assumption for underlying data distribution. The first step is to select the number of k neighbors. The number of nearest neighbors in the kNN algorithm is denoted by K, which is also the deciding factor. If the number of classes is 2, K is generally odd. We get the simplest case of the algorithm when $K = 1$. Then the algorithm is called nearest neighbor. Suppose P1 is the point for which the label must be predicted. In the first step, we need to find the k points nearest to P1 and then classify the points by voting their k neighbors. Each object votes for its class and the class with the most votes is used as the prediction. The closest Points can be found if we use distance measures like Euclidean or Manhattan distance. \cite{mucherino2009k} \cite{navlani_2018}

\textbf{LSTM}

In 1997, LSTM networks were introduced by Hochreiter and Schmidhuber.  LSTM networks belong to the family of recurrent neural networks (RNN). These are the most widely used nets. LSTM networks are capable of learning short-term dependency structures. Their standard behavior is to remember information over long periods of time. Therefore such networks are often used in applications such as classification of time series, generation of sentences, speech recognition, and handwriting recognition. In recent years, the idea of using such networks in forecasting financial time series has become more common \cite{patterson_gibson_2017}. With that being said, they have been increasingly used as a forecasting model. Over time, more LSTM networks have been developed, such as in the work of Z. Li (Financial time series forecasting model based on CEEMDAN and LSTM), which was published in 2018 \cite{cao_li_li_2019}. They improved the forecasting accuracy of various stock indices such as S\&P500 or the DAX by combining the CEEMDAN signal decomposition algorithm with the LSTM model. This should serve as an example for the continuous development of the LSTM networks. A simpler LSTM network was developed through this intuition, which can predict the Brent crude oil price. The prediction is then converted into signals, deciding when to buy and sell.

%% file: 3_Data.tex
\section{Data}\label{Section: Data}

The following chapter aims to analyze the data obtained by the Brent crude oil future data set by the following topics: data description, Brent crude oil, crude oil price determination, financial indicators, log-returns, trading signals, and descriptive statistics. By looking at these various topics, one can fully understand and grasp the full understanding that goes behind the data set.

\subsection{Data description}

For this experiment, one used the daily data of the Brend-Crude-Oil future from 30.07.2007 to 26.04.22. The data used for the experiment was obtained by Yahoo Finance, which can be accessed directly with the yfinance package on Python. The format of the data is called pandas.core.frameData Frame, which is a tabular form of the data. When accessing the data, the following price values were downloaded: Open, High, Low, Close, Adj Close, and the volume. It was found that the price values of a financial instrument show very similar patterns. One first had to process the data and extend it with financial indicators, such as RSI, MACD, Price Rate Of Change, and others that will be discussed later on. Many machine learning methods are classification methods; for such methods, we must create a signal that can be interpreted as a trading signal.

\subsubsection{Brent Crude Oil}

Brent oil is a type of oil that consists of four various fields known as Brent, Forties, Oseberg, and Ekofisk. Brent oil has a limited production of 170’000 barrels per day, which makes up only 0.2 percent of the current worldwide consumption. These barrels will be transported via a subsea pipeline to the Sullom Voe oil terminal on Mainland Shetland, together with the oil from the Ninian field to the final destination of a tanker. Prior to the completion of the pipeline and oil terminal, oil will be loaded onto tankers in the North Sea from loading platforms such as the Brent Spar. Norway’s Statfjord field is found on the East and is connected by two fields by a transfer pipeline. The Brent and Ninian oil fields have now exceeded their production peaks. The traded product is Brent Blend, which is originally a blend derived mainly from the Brent and Ninian fields. The product comes from the North  Sea between the Shetland Islands and Norway. The Brent oil field is being developed by Shell UK Ltd. 

Brent is arguably the most important type of crude oil for Europe due to its low-boiling and light oil texture, which allows it to be more valuable than other crude oils. In addition, it also has a lower sulfur content than other crude oils. It was discovered in 1971 and has been produced since 1976, in the Brent South subfield. It is traded on the ICE Futures commodity exchange in London. \cite{su_2011}

\paragraph{Crude oil price determination}

How is the price of crude oil determined, and how does crude oil trade? The relation of supply and demand determines crude oil prices. When decisions are made because of speculation about possible future developments, psychology plays a crucial role in the stock market. Crude oil prices depend heavily on how traders assess the short- and medium-term direction of the sales market. Fear of political unrest can manifest itself in a price premium that does not in the least reflect real supply conditions. Other actual or expected changes in supply or demand are also reflected in the prices offered on the exchange. The onset of the "driving season" in the spring in the United States of America also makes itself feel like an exceptionally frosty winter in Europe. Incredibly high or low stock levels in important consumer countries lead to decreased or increased product prices.
Moreover, of course, the dollar exchange rate immediately impacts the oil price. Oil barrels or tank farms will not be found on these trading floors. In principle, it is possible to deliver or collect the traded commodity, but this is rarely done in practice. Crude oil is traded on the stock exchange mainly in futures contracts. A futures contract contains a formal, legally binding commitment to buy or sell a predefined quantity of a commodity at a specific time in the future (hence the term "futures"). These very contracts are standardized to ensure that all market participants assume the same qualities, quantities, and delivery terms. As a result, the traded oil futures include only a limited range of products: Crude oil, gasoline, diesel, and heating oil. The exchanges primarily hedge against future price fluctuations for traders when dealing with tangible, "physical" commodities. This strategy is called hedging. It is exciting for oil traders, producers, refiners, and large consumers such as airlines. All these companies have to buy or sell large quantities of petroleum products several months in advance. They are therefore highly exposed to the risk of price fluctuations. Hedgers are not looking for short-term profits on the stock exchange but instead try to balance their positions in the market for the physical commodity and better distribute their risk situations. The majority of the stock market participants pursue a different goal: they try to make short-term gains because they estimate the future market development. These speculators help the market to greater liquidity and take over the dangerous situations that the hedgers try to pass on. However, they can also contribute to the amplification and acceleration of price fluctuations, especially if they are not familiar with the business and technical environment of the oil industry. Thanks to centralized, open business transactions with standardized contracts on the exchange, the prices paid there are exceptionally transparent. \cite{erdoel_vereinigung_2005}

\subsubsection{Financial indicators}\label{financ_indi}

There are various financial indicators that can help investors to make buy or sell decisions. The most important financial indicators are KPIs which belong to fundamental analysis and technical analysis. 

KPIs are used to register how successful specific actions in companies are. All processes in companies can be monitored based on those key performance indicators. With the help of key performance indicators, management and controlling can gain knowledge and analyze processes in companies. With the help of this consistent monitoring, stakeholders can faithfully adapt and improve processes and measures. There is a significant number of selected KPIs. Depending on the company and the area, different KPIs are relevant for measuring performance. For example, marketing focuses on different KPIs than sales, accounting, or management. Which KPIs are the correct ones depends continuously on which actions are to be checked. For example, the KPIs used to study customers' behavior play an exceptionally fundamental role in marketing. On the other hand, management wants to benefit from the project or department performance KPIs. \cite{luber_2019}

In distinction to fundamental analysis, which deals with essential company data, chart analysis is often called technical analysis. This is because it is not the evaluation of business reports in the foreground. Nevertheless, the interpretation of "technical" aids such as chart progressions or indicators of the trends, oscillations, or patterns in the time series that are to  be uncovered. Technical analysis is thus a tool, a method for predicting the likely upcoming price movements of a financial asset - such as a stock or a currency pair - based on market data. Behind the validity of technical elicitation is the demonstration that market participants' collective activities (buying and selling) specifically reflect all relevant data about a traded asset and, therefore, continually assign it a fair market value. In other words, technical traders believe that the market's current or past price action under consideration is the most reliable indicator of future prices. Thus, the technical study can be used for any financial instrument with historic price data. Therefore, chart analysts focus on patterns of price movements, trading signals, and various other analytical charting tools to evaluate the strength or weakness of an asset. All chart analysis product packages have one thing in common: the assumption that observable occurrences repeat with the probability that can be predicted within limits. It becomes arguable to use specific patterns in charts or the values of indicators as a guide for the trader. Technical analysis is not only used by chart analysts. Plenty of investors use fundamental analysis to determine whether they should buy into a market or not. Nevertheless, after this decision is made, they also use chart analysis to register excellent entry price levels. \cite{gehrt_2020}

\paragraph{Returns and Log Returns}

Rates of return and logarithmic rates of return are the net percentage gains or losses of the past cost of an investment over a specified period. They calculate the percentage rate of change over a defined period such as a day, month, or year. Returns can be used to measure the increase in value of any asset, including stocks, bonds, mutual funds, real estate, collectibles, etc. Investors can also use the returns to compare past periods or other investments. Two inputs are required to calculate the rate of return: i) the purchase amount of the investment and ii) the current value or terminal value of the investment in the period being measured. In some cases, any income generated from the asset such as interest and dividends is also included in the calculation. A rate of return is calculated by subtracting the initial value of the investment from its current value and then dividing it by the initial value. To express the return in percent, the result is multiplied by 100. A simple logarithmized rate of return is calculated by taking the quotient of the initial value of the investment by the current value of the investment and logarithmizing. To express the return in percent, multiply the result by 100. \cite{best_2022} \cite{quantivity_2011}

\begin{equation}
   \text{Returns in percent} = \frac{r_{t} - r_{t-1}}{r_{t-1}} * 100
\end{equation}

\begin{equation}
   \text{Logarithmic Returns in percent} = log(\frac{r_{t}}{r_{t-1}}) * 100
\end{equation}

\paragraph{RSI}

The Relative Strength Index or RSI is a momentum trading indicator often used in technical analysis. The RSI can take values between 0 and 100 and display them as an oscillator. It measures the impact of the price change in a given period. The default period is 14 days but can be changed to any. The RSI is used to measure overbought or oversold conditions in a stock or other asset price. The usual interpretation of the RSI is that values greater than or equal to 70 indicate an overbought condition (overvaluation of the asset) and will cause the price to fall. Values less than or equal to 30 indicate an oversold condition (undervaluation of the asset), and a price slope follows.\cite{admirals_2022}  The RSI is calculated in the following three steps: 

First, the sum of all positive and negative price changes is calculated:

\begin{equation}
    \text{sum}_\text{up}(t) := \sum_{i=1}^{n} \max \left\{ P(t-i+1) - P(t-i); 0 \right\}
\end{equation}
\begin{equation}
    \text{sum}_\text{down}(t) := - \sum_{i=1}^{n} \min \left\{ P(t-i+1) - P(t-i); 0 \right\}
\end{equation}
%\left\{  \right\}
Then the mean value of the sums is taken:
\begin{equation}
    \text{avg}_\text{up}  := \frac{\text{sum}_\text{up}} {n}
\end{equation}
\begin{equation}
    \text{avg}_\text{down} := \frac{\text{sum}_\text{down}}{n}
\end{equation}

The RSI then results with:
\begin{equation}
    \text{RSI} := 100-(\frac{100}{1-(\frac{\text{avg}_\text{up}}{\text{avg}_\text{down}})})
\end{equation}

\paragraph{Stochastic Oscillator}

The stochastic oscillator, also known as the stochastic indicator, is a popular trading helpful indicator for predicting trend reversals. The indicator works by focusing on the location of an instrument's closing price relative to the high-low range of the price over a given period. Typically, 14 prior periods are used. By comparing the closing price to previous price movements, the indicator predicts price reversal points. The stochastic indicator is a two-line indicator applied to any chart. It fluctuates between 0 and 100. The indicator shows how the current price behaves compared to the highest and lowest price levels in a predefined period. If the stochastic indicator is at a high level, the instrument's price has closed near the upper end of the period range. If the indicator is at a low level, the price has closed near the lower end of the period range. The general rule for the stochastic indicator is that prices in an upmarket close near the high. In contrast, in a down market, prices close near the low. When the closing price deviates from the high or low, it indicates a slowdown in momentum. The formula of the Stochastic Oscillator is as follows: \cite{hayes_2021}  
\begin{equation*}
\begin{split}
    C     & = Current \ Close \ Price  \\ 
    H_{x} & = The \ Highest\ High\ value\ of\ the\ last\ x\ periods \\
    L_{x} & = The\ Lowest\ Low\ value\ of\ the\ last\ x\ periods\\
\end{split}
\end{equation*}
\begin{equation}
    Percent_K = (\frac{C - L_{x}}{H_{x} - L_{x}}) * 100
\end{equation}

The idea behind the stochastic indicator is that the dynamics of the price of an instrument often change before the instrument's price movement changes direction. It also focuses on price momentum and can be used to identify overbought and oversold levels in stocks, indexes, currencies, and many other assets.

\paragraph{MACD}

The moving average convergence divergence (MACD) indicator calculates the change or speed of price movement of the asset. It is a momentum indicator that follows a trend and shows the correlation between two moving averages of the asset.
The formula for calculating the MACD is simple. MACD is a subtraction of the 26-period exponential moving average (EMA) from the 12-period EMA.

\begin{equation}
    MACD = \mathrm{EMA_{12}}\ -\ EMA_{26} \\
\end{equation}

EMA is the moving average that gives more weight to the current price points. This allows this moving average to be more responsive to recent price changes.\cite{guertel_2022} 

\paragraph{Price Rate Of Change}

The Price Rate of Change or (ROC) conveys exactly the same statements as the "Momentum". Momentum measures the speed, force or strength of a price movement by subtracting the closing price n periods ago from today's closing price. The difference between the two indicators is simply that in the ROC the difference calculated is divided by the closing price n periods ago and this quotient is then multiplied by 100. The current ROC value thus shows by how many percent today's price is above or below the price n periods ago. Some programs omit the multiplication by 100, which means that the results are not displayed at a 100 percent line, but at the zero line. This is purely program-related, the statement is the same. The ROC shows the momentum of the price movement. Accordingly, the position of the ROC line is important. (ROC in the positive area) A rising ROC indicates an (increasing) positive momentum, i.e. a continuation of the upward trend. A falling ROC indicates a decrease in momentum and thus a possible end of the upward movement. (ROC in negative range) A falling ROC indicates a (increasing) negative momentum, thus a continuation of the downward trend. A rising ROC indicates an easing of the negative momentum and thus a possible end of the downward movement. Thus, if recent price gains are less than previous ones, the ROC will fall in positive territory, and analogously, if recent price declines are less than previous ones, the ROC will rise in negative territory. Thus, the ROC can turn even though the stock price is forming new highs or lows. The ROC can be calculated as follows \cite{elearnmarkets_2022}: 

The Price Rate of Change (ROC) conveys the same statements as the "Momentum." Momentum measures a price movement's speed, force, or strength by subtracting the closing price n periods ago from today's closing price. The difference between the two indicators is simply that in the ROC, the difference calculated is divided by the closing price n periods ago. This quotient is then multiplied by 100. The current ROC value thus shows by how many percent today's price is above or below the price n periods ago. Some programs omit the multiplication by 100, which means that the results are not displayed at a 100 percent line, but at the zero line. This is purely program-related; the statement is the same. The ROC shows the momentum of the price movement. Accordingly, the position of the ROC line is essential. (ROC in the positive area) A rising ROC indicates an (increasing) positive momentum, i.e., upward trend continuation. A falling ROC indicates a decrease in momentum and thus a possible end of the upward movement. (ROC in negative range) A falling ROC indicates an (increasing) negative momentum, thus continuing the downward trend. A rising ROC indicates an easing of the negative momentum and thus a possible end of the downward movement. Thus, if recent price gains are less than previous ones, the ROC will fall in positive territory, and vice versa, if recent price declines are less volatile than previous ones, the ROC will rise in negative territory. The ROC can turn even though the stock price forms new highs or lows. The ROC can be calculated as follows \cite{elearnmarkets_2022}: 

\begin{equation*}
\begin{split}
    n       & = \mathrm{Number}\ of\ periods \\ 
    C_{t}   & = \mathrm{Current}\ close\ price\\
    C_{t-n} & = \mathrm{Close}\ price\ n\ periods\ before\\
\end{split}
\end{equation*}

\begin{equation}
    ROC_{t} =\frac{C_{t} - C_{t-n} }{C_{t-n}} * 100
\end{equation}

\paragraph{Trading Signal}

A user can freely define the trading signal. The returns are used for coding signals (binary signals) in this work. Positive returns are coded as the number one, and negative returns are coded as zero. After encoding the signal, the signal is shifted back to one line to realize a one-day prediction. This was done in the context of classification algorithms like kNN.

\begin{equation}
    Signal_{t} = \left\{ \begin{matrix}
1  & \mathrm{,if}\ return_{t+1} > 0 \\
0  & \mathrm{,if}\ return_{t+1} <= 0
\end{matrix} \right.
\end{equation}

\subsection{Descriptive statistics}

The Brent Crude Oil dataset was extended with the financial indicators, and all irrelevant values were filtered out. After that, all rows with NA values were deleted to clean up the dataset. After editing, the dataset consists of the following columns: Close Price, Log Returns, RSI, MACD, K Percent, ROC, and Signal. A closer look at the data will show various outcomes such as how the data is distributed, which are the most critical parameters, which columns contain outliers, and can already be made certain statements about how the data will behave in the future.

\newpage 

\subsubsection{Time Series Brent Crude Oil}

\begin{figure}[ht]
\includegraphics[width=.5\linewidth]{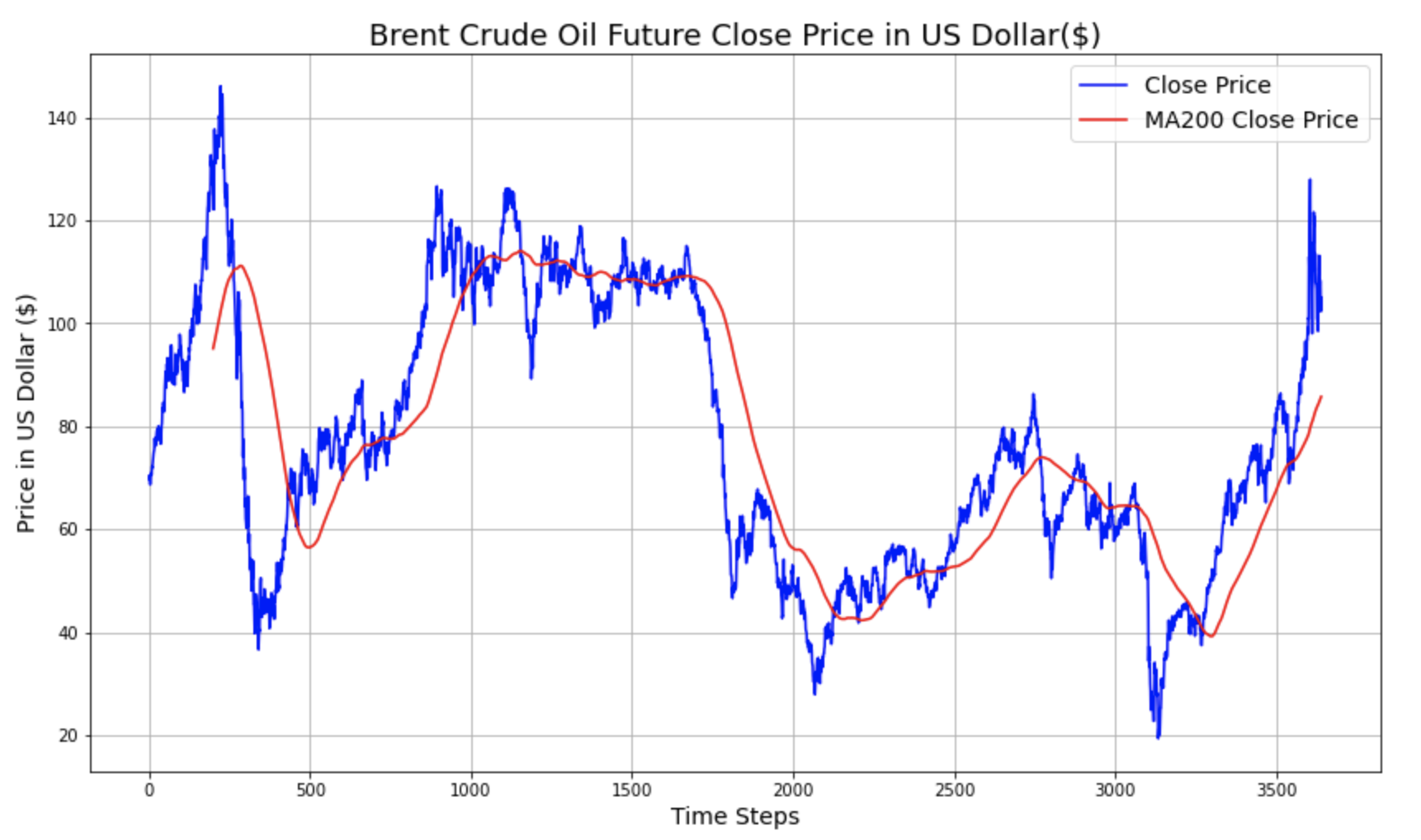}
\hfill
\includegraphics[width=.5\linewidth]{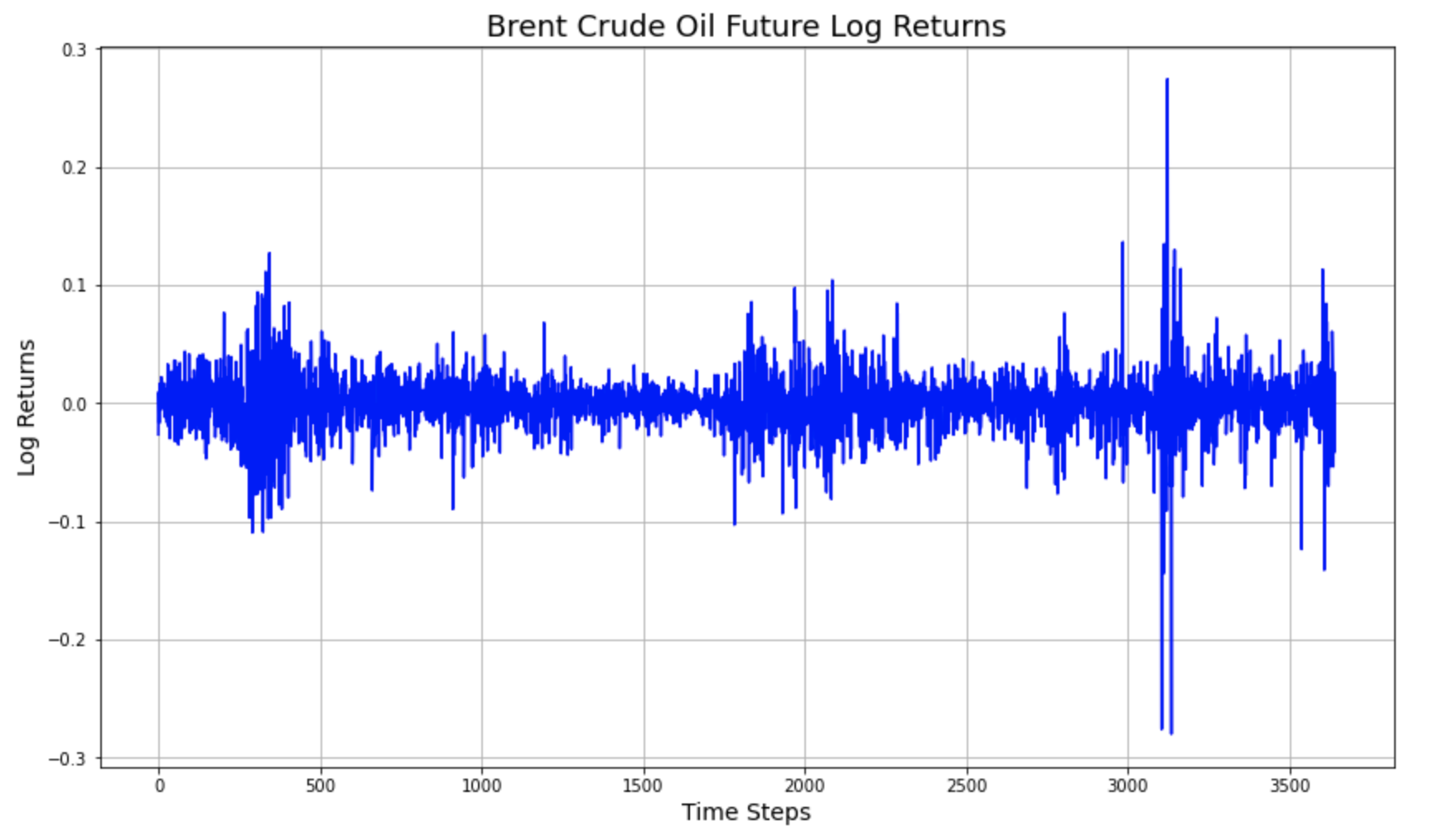}
\caption{Close Price from Brent Crude Oil Future (left) and Log Returns from Brent Crude Oil Future (right)}
\label{Fig1}
\end{figure}

In periods of high volatility, the prices can change very rapidly, and the prices also tend to fall in such cases. The time series has only shown a clear upward or downward trend in specific time intervals in recent years. In figure \ref{Fig1}, the Brent Crude Oil feature closing price with a simple moving average over 200 days, and the Log Returns can be seen.

As depicted in table \ref{Tab1}, the essential position measures are shown. The average closing price is USD 77.4, as shown in figure \ref{Fig1}. The average of the log-returns is more interesting because it is a positive number, which means that the Brent Crude Oil Future gives a positive return on average, which is a good measure for investments. In addition, the quantiles shows that the chance of making a profit is more remarkable than making a loss, which provides us with information that the buy-and-hold strategy is profitable.

\begin{table}
\hfill
\begin{centering}
\caption{Position measure (Close Price and Log Returns)}
\begin{tabular}{cccc}
\toprule[0.1pt]
\textbf{Statistic} & \textbf{Close} & \textbf{Log Returns}  \\\addlinespace
\midrule[0.1pt]\addlinespace
Mean & 77.384 & 0.000105 \\\addlinespace
Standard deviation & 25.919 & 0.025 \\\addlinespace
Minimum & 19.330 & -0.280 \\\addlinespace
5\% & 41.290 & -0.037 \\\addlinespace
25\% & 55.980 & -0.010 \\\addlinespace
50\% & 73.360 & 0.0004 \\\addlinespace
75\% & 103.810 & 0.011 \\\addlinespace
95\% & 117.100 & 0.035 \\\addlinespace
Maximum & 146.080 & 0.274\\\addlinespace
\bottomrule[0.1pt]\addlinespace[2pt]
\label{Tab1}
\end{tabular}\par
\end{centering}
\end{table}

\newpage

\begin{figure}[ht]
\begin{centering}
\includegraphics[width=.5\linewidth]{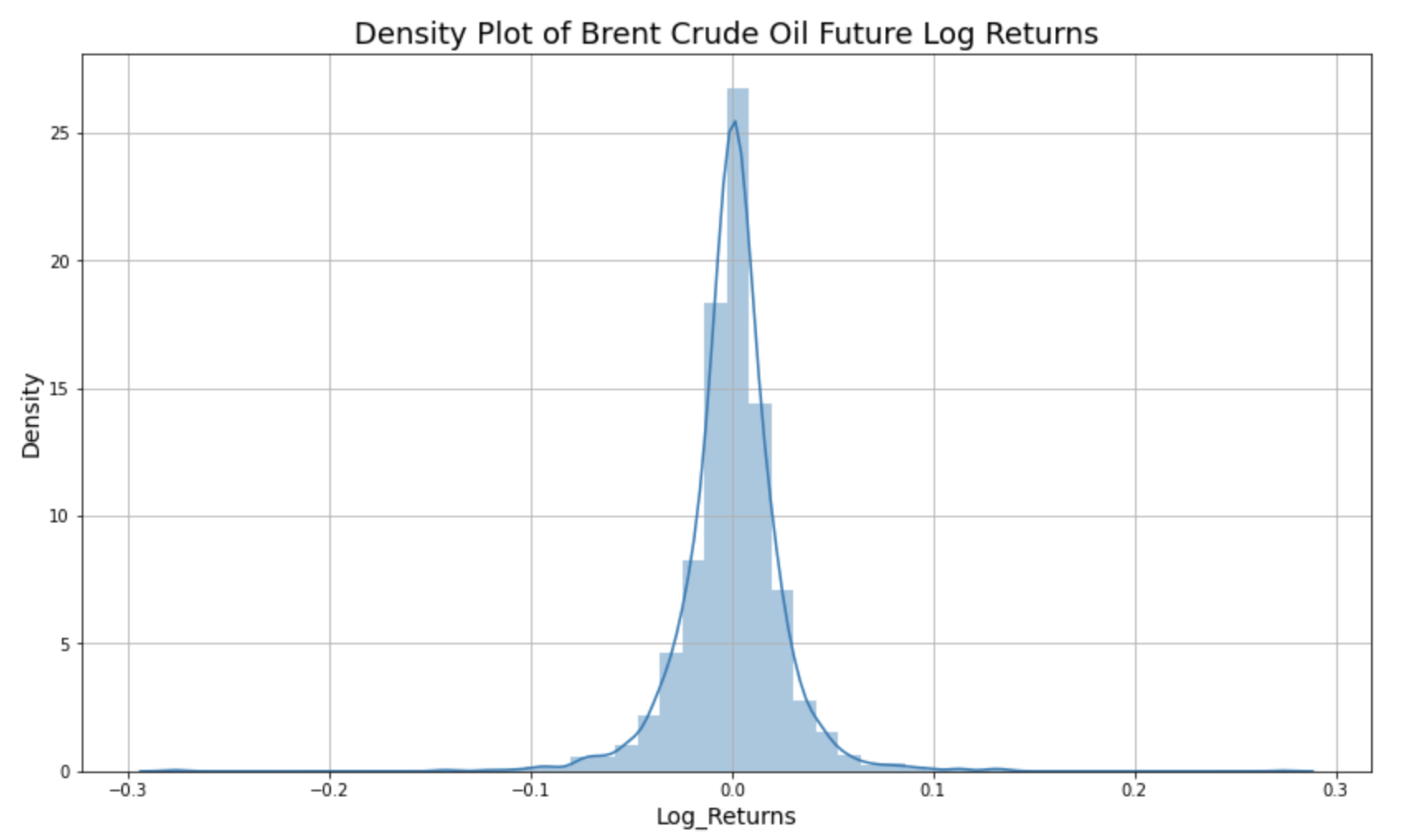}
\caption{Density Log Returns}
\label{Fig2}
\end{centering}
\end{figure}

\begin{figure}[ht]
\includegraphics[width=.5\linewidth]{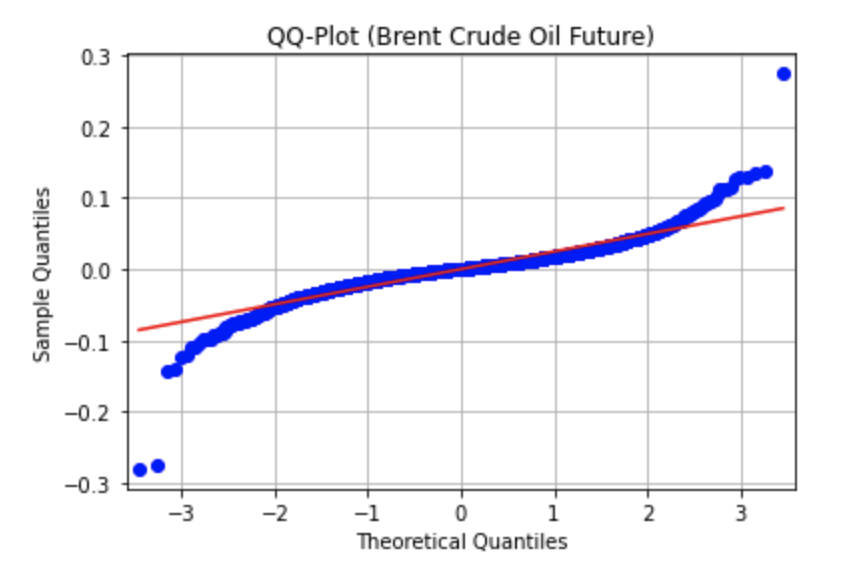}
\includegraphics[width=.5\linewidth]{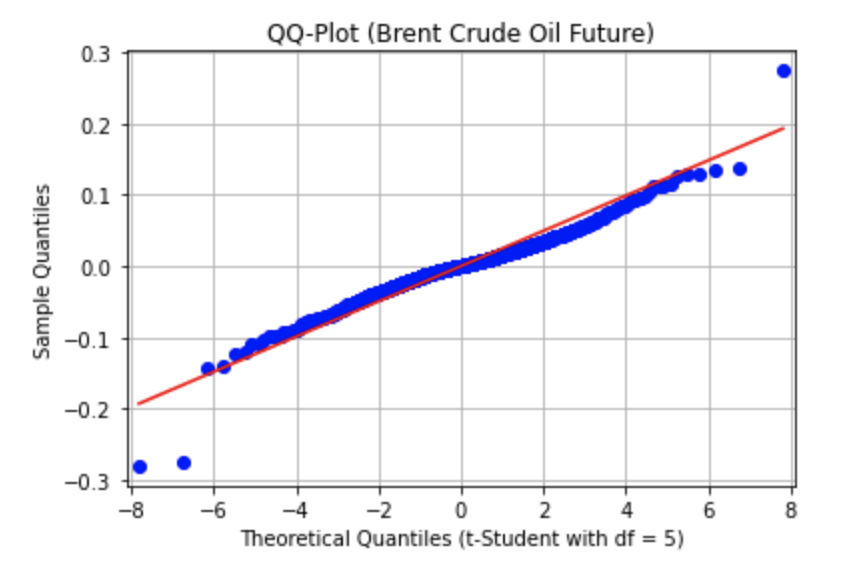}
\caption{Normal QQ-Plots (left) and t-Student QQ-Plot with df = 5 (right)}
\label{Fig3}
\end{figure}

The distribution of the log-returns (figure \ref{Fig2}) shows that the distribution is symmetric around zero. It is visible that the distribution is not normally distributed with the help of a quantile-quantile normal plot (figure \ref{Fig3}); it is possible to verify this. The distribution's tails are immense; for this reason, a t-Students distribution is to be assumed here.

\newpage

\begin{figure}[ht]
\begin{centering}
\includegraphics[width=.5\linewidth]{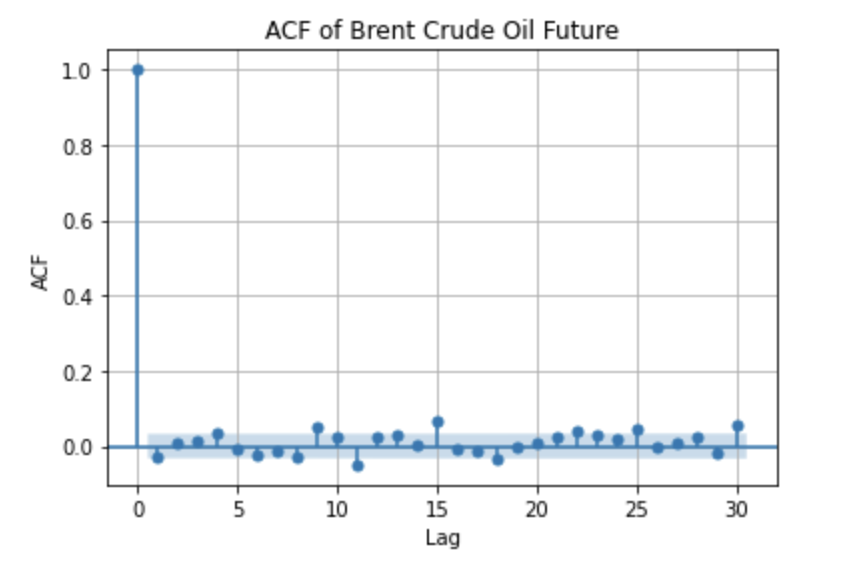}
\caption{Autocorrelation Plot}
\label{Fig4}
\end{centering}
\end{figure}

To check this, another quantile-quantile plot was used. But this time with the distribution assumption of a t-Students distribution with 5 degrees of freedom. See figure \ref{Fig3}.
It is obvious that the points fit better around the red line. The 3 points that are not on the red line can be considered as outliers.
The autocorrelation of the log returns look like this figure \ref{Fig4}.

The distribution assumption of a t-Students distribution was used with 5 degrees of freedom (see figure \ref{Fig3}). The points fit better around the red line. The 3 points which are not on the red line are outliers. The autocorrelation of the log-returns looks like in figure \ref{Fig4}. The dependency structure is shallow.

\begin{figure}[ht]
\includegraphics[width=.5\linewidth]{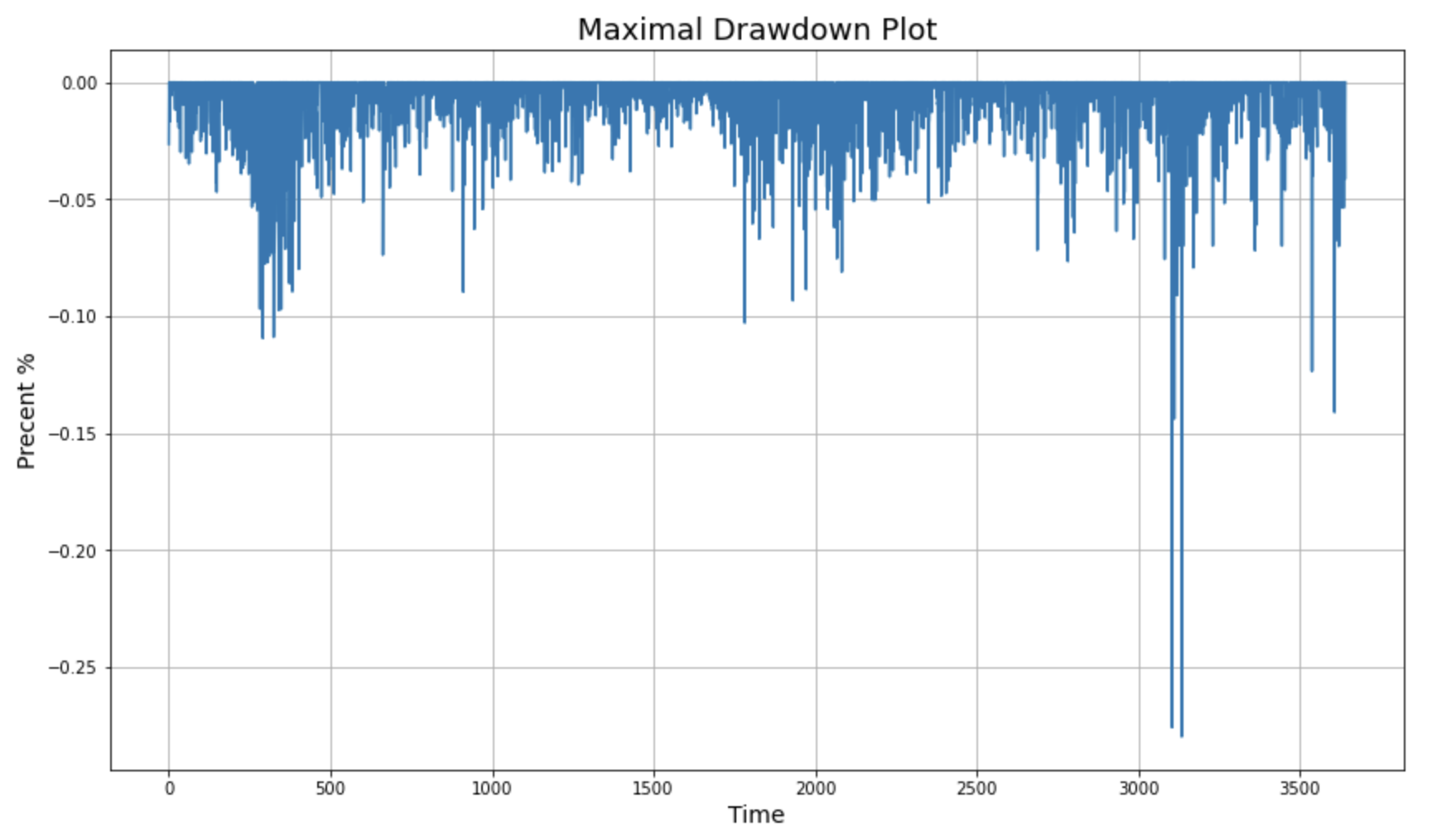}\hfill
\includegraphics[width=.5\linewidth]{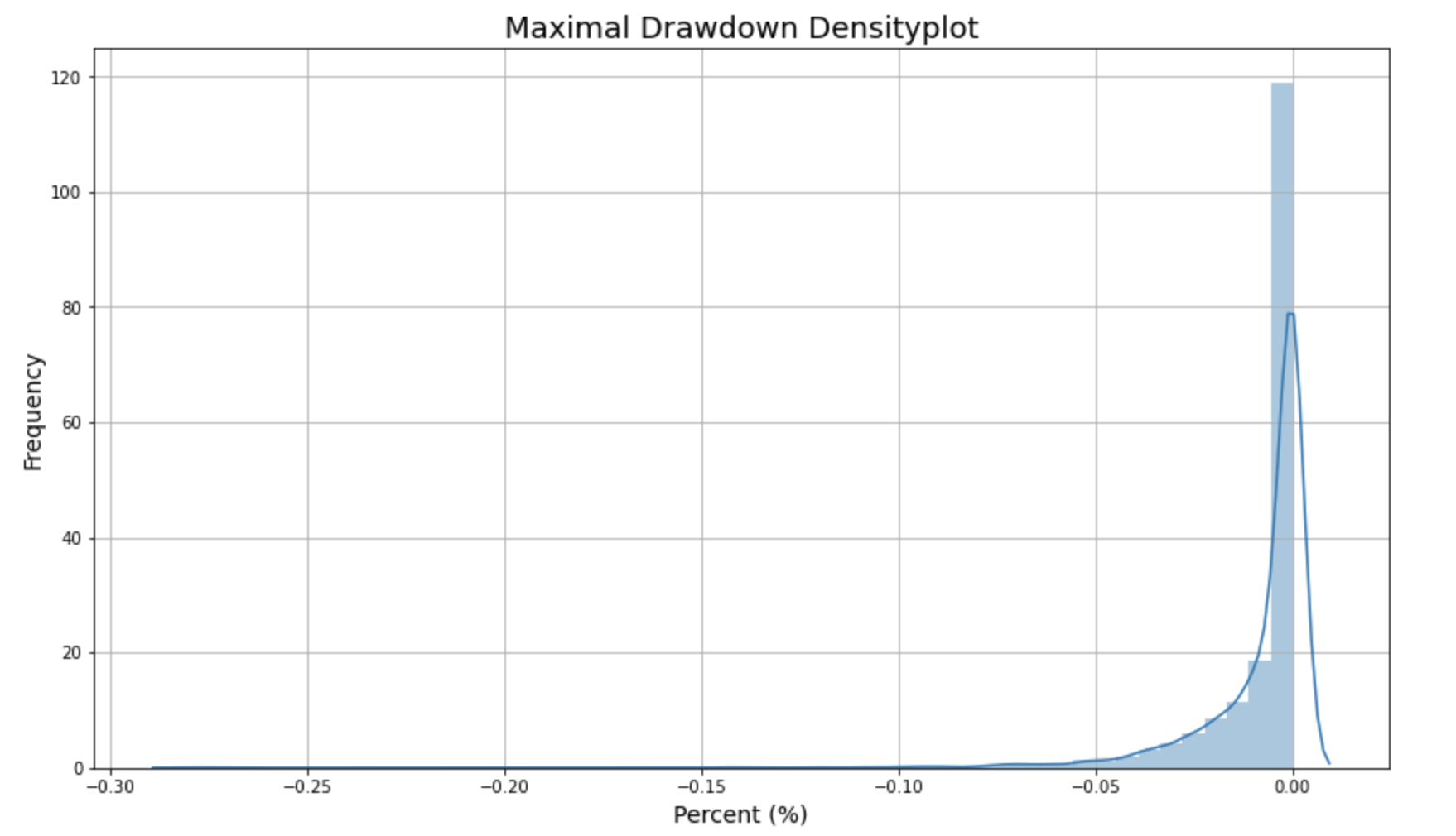}
\caption{MDD Plot (left) and MDD Density Plot (right)}
\label{Fig5}
\end{figure}

The maximum drawdown (MDD) is a measure of the risk of an investment fund. It represents the maximum cumulative loss from the highest price within a period under consideration and is usually presented as a percentage value. Figure \ref{Fig5} shows the MDD of the whole available data and its distribution.

\newpage

Table \ref{Tab2} is an overview of the most important parameters of the total MDD. The quantiles show that no losses are generated in most cases, which confirms the assumption made based on the table (Log Returns). It can be seen (except for a few outliers) that the maximum losses are around -0.1 percent and that the losses are usually minimal.

\begin{table}
\hfill
\begin{centering}
\caption{Maximum Drawdown Position measure}
\begin{tabular}{cccc}
\toprule[0.1pt]
\textbf{Statistic} & \textbf{MDD}   \\\addlinespace
\midrule[0.1pt]\addlinespace
Observations & 3641\\\addlinespace
Mean & -0.008 \\\addlinespace
Standard deviation & 0.016 \\\addlinespace
Minimum & -0.280 \\\addlinespace
5\% & -0.037 \\\addlinespace
25\% & -0.010 \\\addlinespace
50\% & 0 \\\addlinespace
75\% & 0 \\\addlinespace
95\% & 0 \\\addlinespace
Maximum & 0 \\\addlinespace
\bottomrule[0.1pt]\addlinespace[2pt]
\label{Tab2}
\end{tabular}\par
\end{centering}
\end{table}

\begin{figure}[ht]
\includegraphics[width=.48\linewidth]{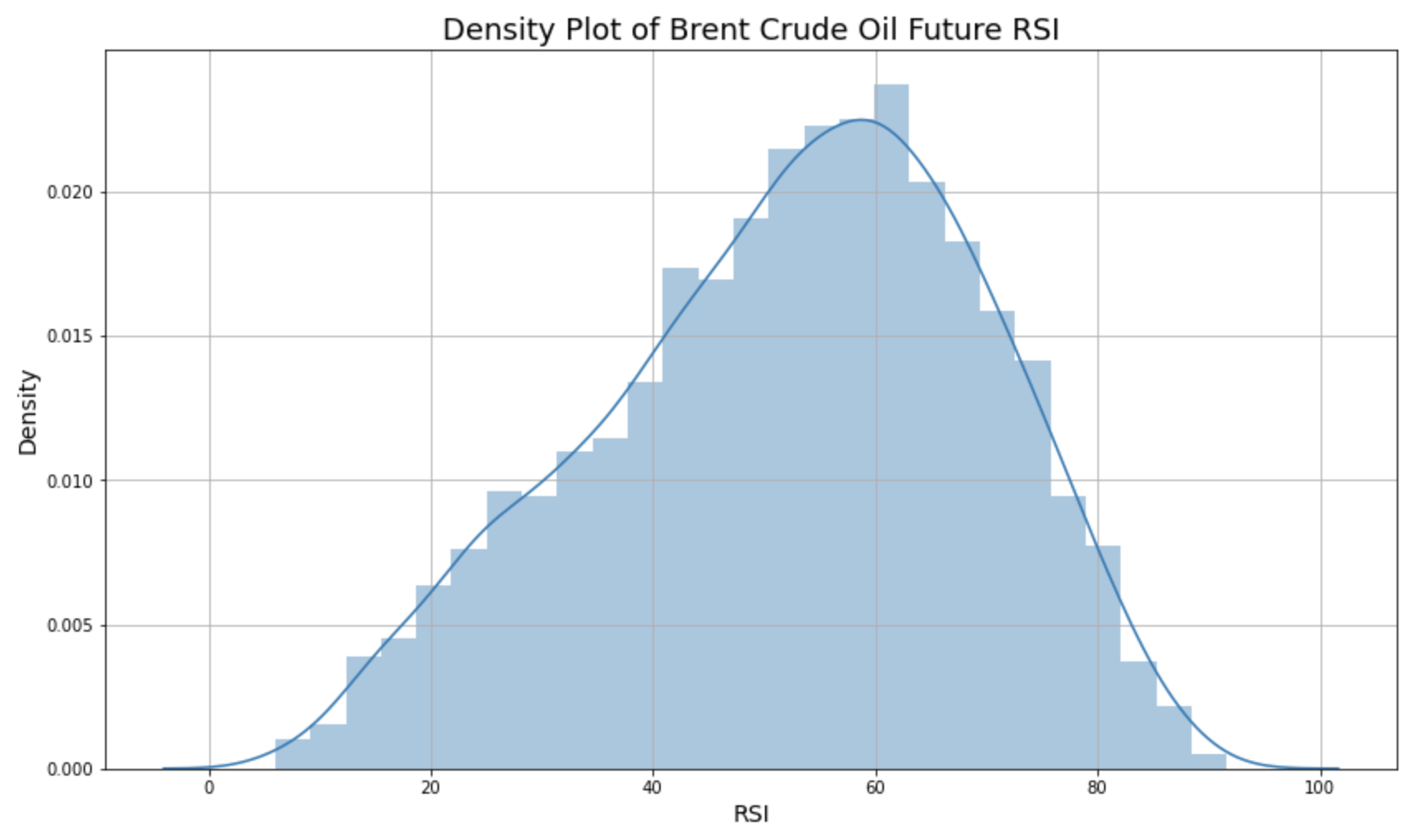}\hfill
\includegraphics[width=.48\linewidth]{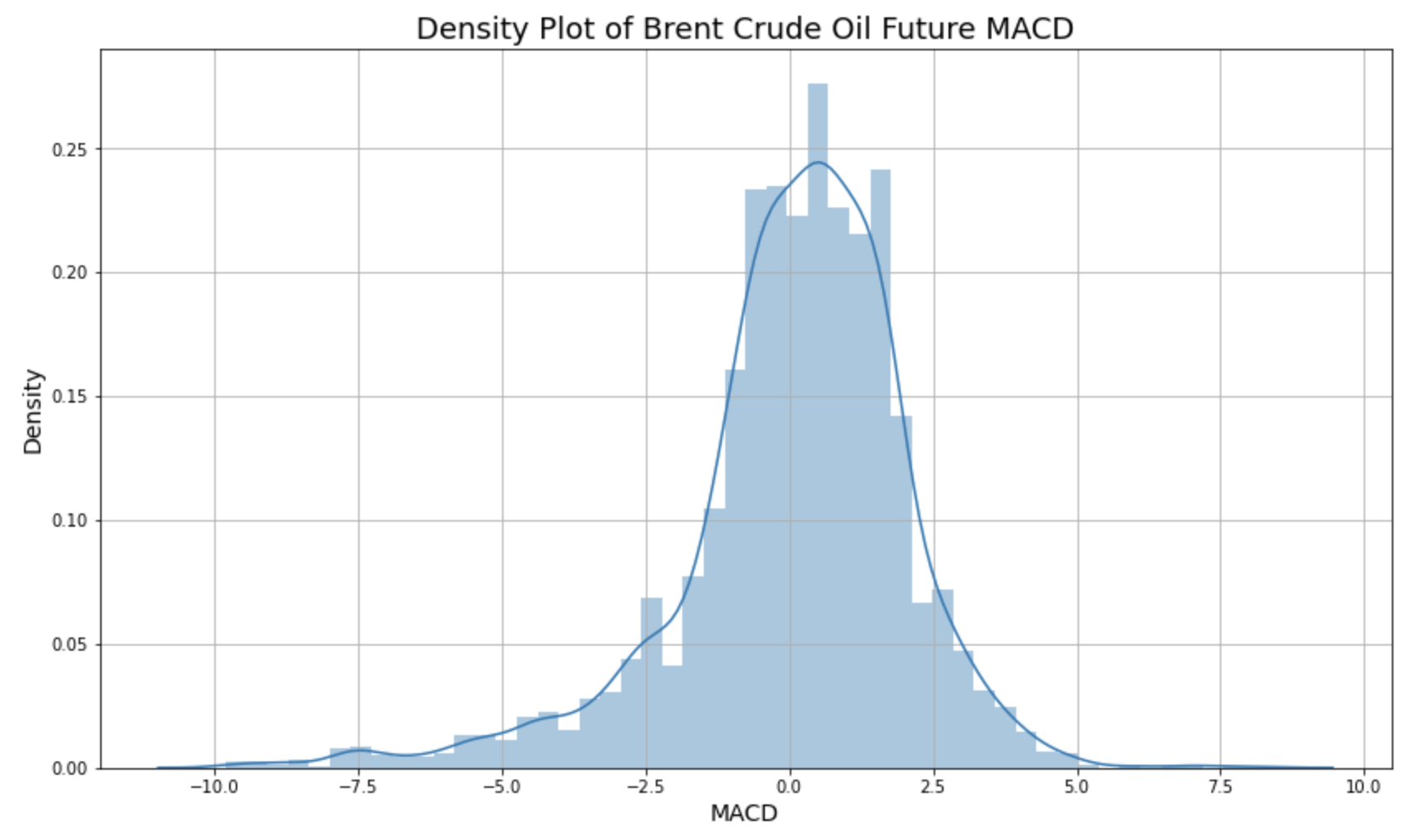}\hfill
\includegraphics[width=.48\linewidth]{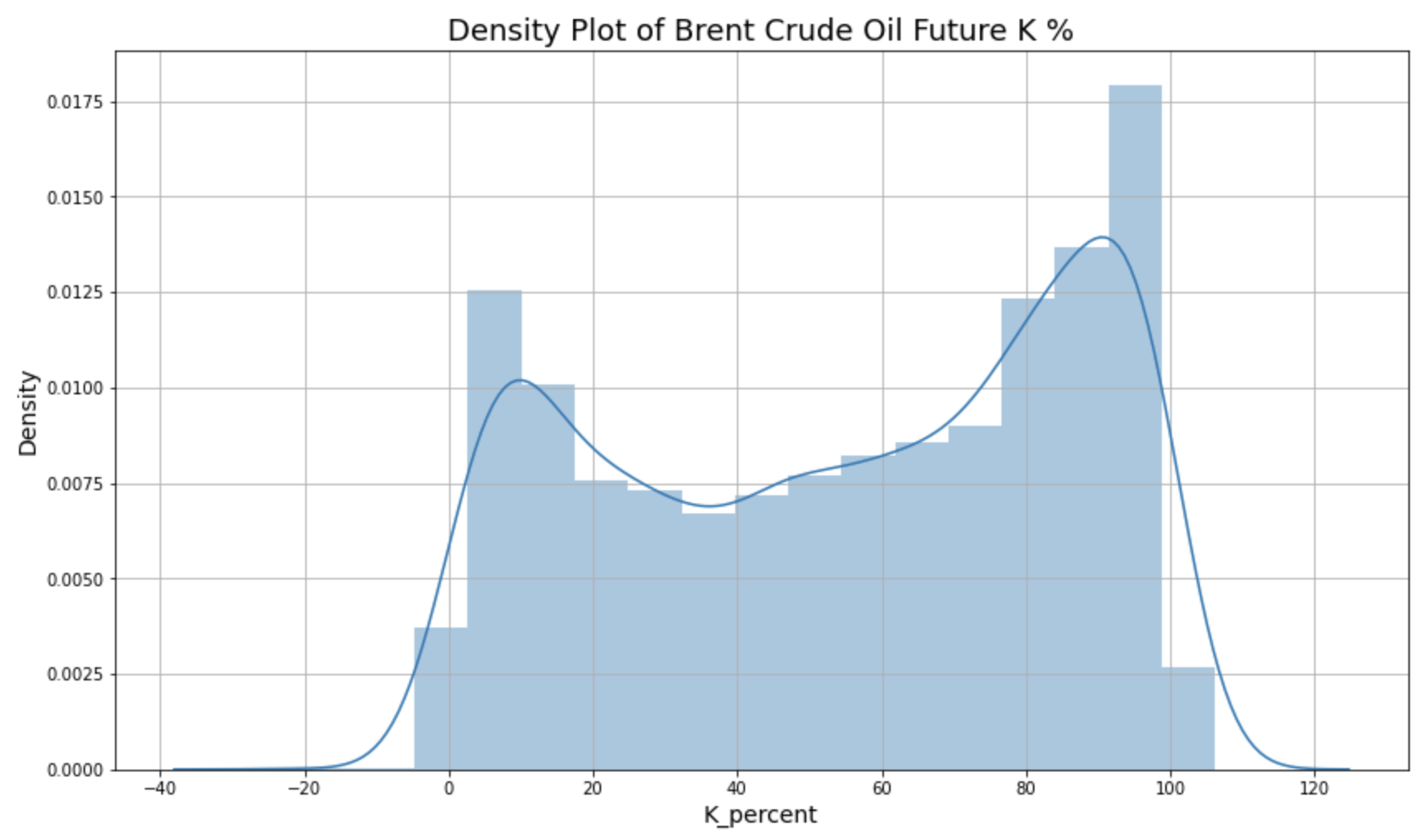}\hfill
\includegraphics[width=.48\linewidth]{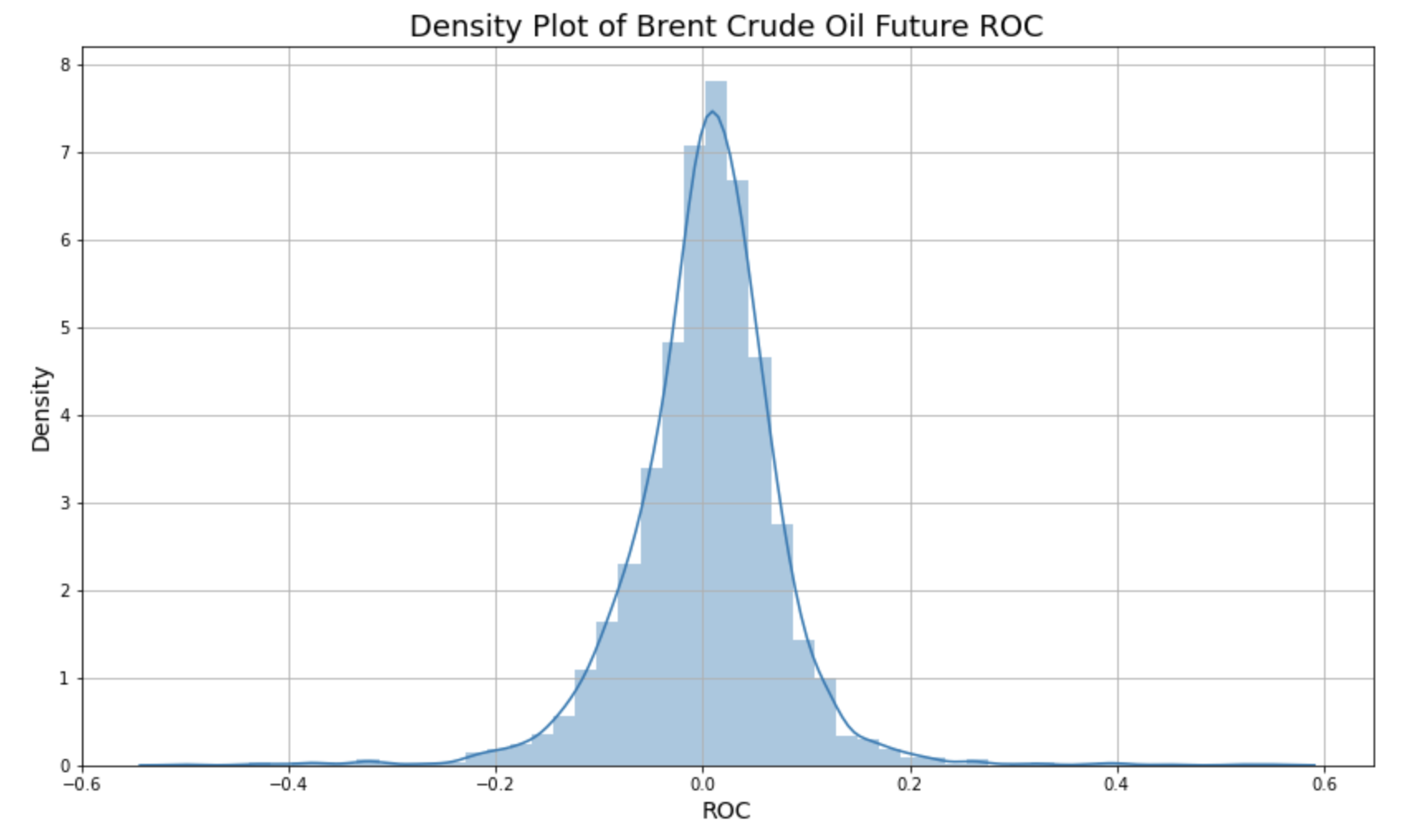}\hfill
\caption{Feature Density Plots}
\label{Fig6}
\end{figure}

The distributions of the features in figure \ref{Fig6}  have a bell shape with the exception of the K percent. This feature is a bimodal distribution, and depending on the Machine Learning algorithm, it is advisable to transform the data to be as normally distributed as possible.

%% file: 4_Experiment.tex
\section{Experiment}\label{Section: Experiment}

Within this chapter, there were six different models presented that had different applications and results. For each model, two possible strategies were simulated. The first strategy focuses on a constant long strategy and the second strategy focuses on a long-short strategy. In the long strategy, a signal vector was generated, coded with 1 and 0. Code 1 is a buy signal, and code 0 is a sell signal. In this strategy, either future asset was bought or sold. Moving onto the long-short strategy that was coded with the signal 1 or -1, this strategy allows short selling. Short-selling strategies have a higher risk appetite but can yield significantly higher returns when investing accurately. It is important to note that both strategies mentioned above were compared with the buy and hold. Additionally, two other models from the classical time series analysis and chart analysis were used to compare the machine learning models, the ARMA-GARCH, and cross signal, models. The ARMA-GARCH model belongs to the classical time series models, while the cross signal strategy is found in the chart trader. All models were experimented and tested with the same data to enable a complete and correct comparison, not excluding the prediction horizon is harmonious for all models. Financial indicators are used individually or combined in the chart analysis to develop a trading strategy by using trading rules that include these financial indicators. The theory behind the machine learning models is to give exact financial indicators in the hope that the models find patterns in the data to predict the up and down signal for the next day. By looking at \ref{Tab3}, one can find all the input variables used for all machine learning models. The first four input variables are features used uniformly for all ML models. The label “signal“ was used for classification and Close Price for regression. The labels were each shifted back one row to allow for one-day prediction. Since the values, “features” should predict tomorrow's value of the label $Label_{t+1}$.

\begin{table}
\hfill
\begin{centering}
\caption{Description of input variable}
\begin{tabular}{cccc}
\toprule[0.1pt]
\textbf{S. no.} & \textbf{Input variable formulation} & \textbf{Type} \\\addlinespace
\midrule[0.1pt]\addlinespace
1 & RSI & Feature\\\addlinespace
2 & ROC & Feature\\\addlinespace
3 & MACD & Feature\\\addlinespace
4 & K-percent & Feature\\\addlinespace
5 & Signal & Label\\\addlinespace
6 & Close Price & Label \\\addlinespace
\bottomrule[0.1pt]\addlinespace[2pt]
\label{Tab3}
\end{tabular}\par
\end{centering}
\end{table}

Other input data were used for the LSTM, ARMA-GARCH, and the Cross Signal Model. This is explained in more detail in the respective section.

\subsection{Architecture of the Algorithms}

\begin{figure}[ht]
\begin{centering}
\includegraphics[width=17cm]{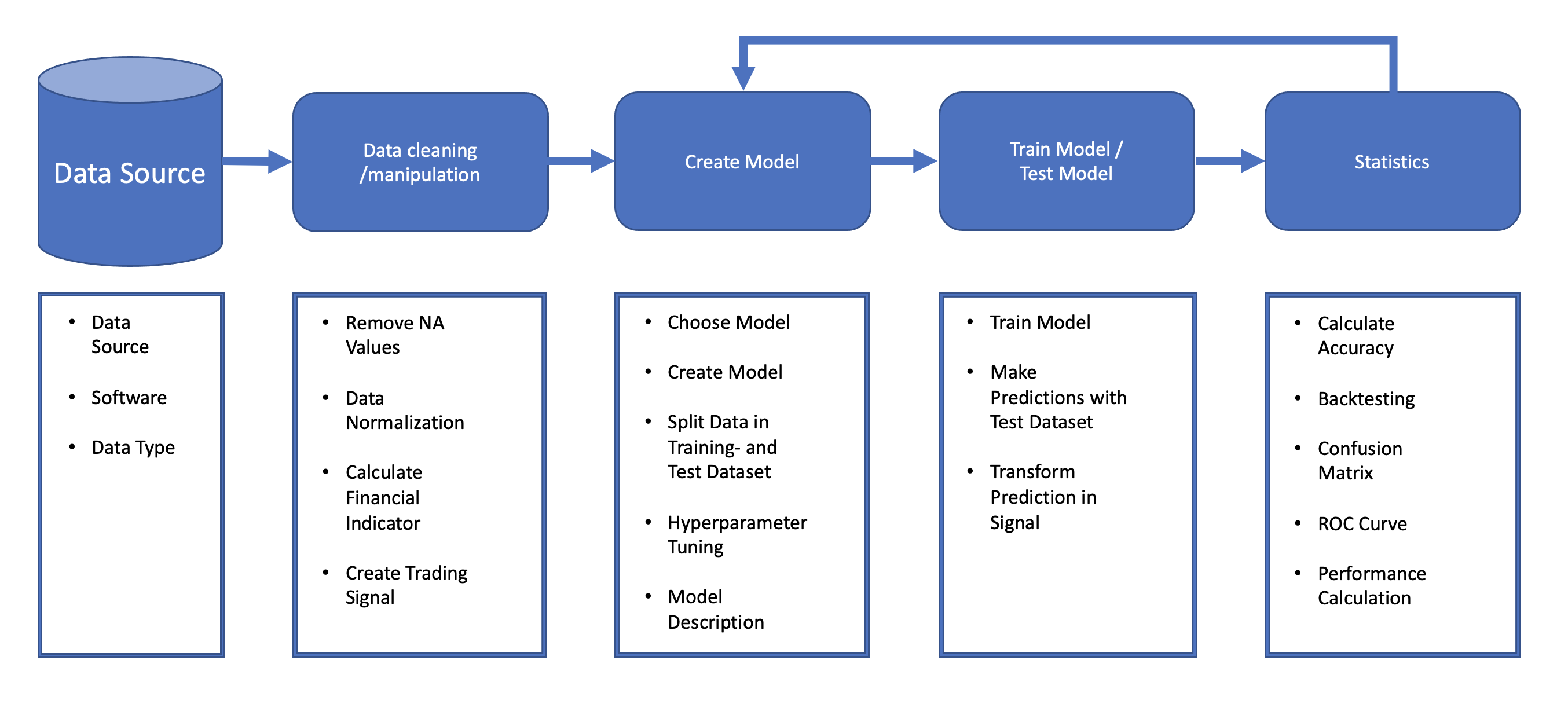}
\caption{Architecture}
\label{Fig7}
\end{centering}
\end{figure}

The architecture of the algorithms used in this work is shown in figure \ref{Fig7}. The process flow for all ML algorithms as well as for the traditional time series methods can be divided into five categories. First comes the data acquisition, where considerations must be made such as what software will be used to process the data, where will the data come from, and what type of data will be needed. In this work all computations and analyses were compiled with the programming language Python. The data was downloaded directly as a data frame via a package integrated in Python (yfinance).

In the second step comes the data preparation, here it is important to examine the data carefully because often the data are incomplete and filled with NA values. If this is the case it is necessary to check where the affected rows are and delete them if possible. If there are many NA values, it is advisable to repeat the first step and load the data in another database. If this is also not possible, the NA values can be processed with missing-data algorithms. Once the data has been processed so that there are no more NA values, the data can be manipulated and extended with new features. In trading, financial indicators are often used to gain a better understanding of the time series. For certain ML algorithms it is better if the data is normalized. That means you scale the data for example with the minmax scaler. If it is not clear in advance which ML algorithm will be used for the analysis, the normalization can be done later. Because many ML algorithms are classification algorithms it is important to create a feature that has a class. In this work a signal was created as a class that encodes the returns with zero and one. Third, a model is chosen and then created in a programming language. It is advisable to look closely at the functions of each algorithm. Most programming languages have web pages where you can read about how the functions work, what parameters exist and examples of use. With this information you can then select the relevant features and also the target variable also called label. With most algorithms there are functions with which one can realize the hyperparameter tuning simply. To realize hyperparameter tuning, the data set must first be split into a training and a test data set. Since hyperparameter tuning needs the data to find the optimal parameters for the given data set. Once these steps are done, the model can be built. For the fourth step you have to pass the whole training data-set to the model, i.e. all features and the label. With this data the model is trained. Depending on the model this can take some time. After the model is trained, we can make a prediction with the test data-set. The model is passed only the features of the test data-set. The output of the model is then the prediction of the label values. At last comes the fifth step, here we compare the results of the prediction and the actual values. In this work, we have always used the same comparison method, whether regression or classification. Here we had to convert the results of the regression back into a signal. The usual methods are, the calculation of the accuracy, the confusion matrix. If the results of the backtesting are satisfactory, the performance of the different trading strategies can be calculated. If the results are bad, you have to go back to the third step.\cite{lv_yuan_li_xiang_2019} 

\newpage
\subsection{Benchmark Models}
In this chapter, the results of the ARMA-GARCH model and the cross signal strategy are presented. A detailed procedure of how everything was done can be seen in the appendix, in this chapter only the final result are presented.

\subsubsection{ARMA-GARCH}

\begin{figure}[htb]
    \centering
    \begin{minipage}[h]{0.7\linewidth}
        \centering
        \includegraphics[width=0.9\linewidth]{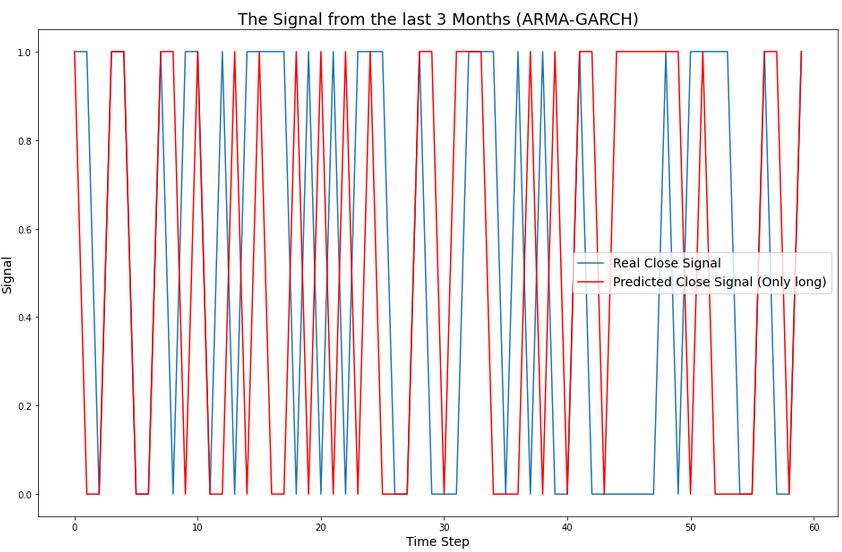}
        \caption{Signals of the last three months}
        \label{Fig8}
    \end{minipage}
\end{figure}

The application of ARCH/GARCH models has become unthinkable in the financial industry. First, a detailed analysis of the financial time series was done, which showed that the data are not normally distributed and that they follow a Students-t distribution. This was supported through a QQ-Plot analysis. The Jarque-Bera test confirms the assumption of no normal distribution. Figure \ref{Fig1} shows the prices and log returns of the Brent crude oil time series. By chart analysis, no significant trend is apparent on average for the time being. And performing the Augmented Dickey-Fuller test revealed no significant evidence of non-stationarity.  Subsequently, the autocorrelation functions ACF and PACF are performed to determine the model order. Here we rely on the Bayesian information criterion BIC. The best model was an ARMA(1,1), which had the lowest BIC. However, if we look at the log returns, we see that there are various volatility clusters, which is due to the heteroscedasticity. With the GARCH(1,1) we could capture the heteroskedasticity well which led to a very good performance. For training and testing the model, log returns from Brent crude oil are used. In the next step, the data were split into 80:20 proportions with which the ARMA(1,1)-GARCH(1,1) model was trained (80\% of data) and the predictions were converted into binary numbers -1 and 1 using the signum operator. With a positive signal you are long, with a negative signal you are short. Thus we get the signals with which the trading performance can be calculated. The signals can be seen in the figure \ref{Fig8}. The model was able to recognize the volatile market phases well. The accuracy of the predicted signals was 54.27\%, which is good. Afterwards a detailed extreme value analysis was don, to see how the model have performed in high volatility phases and in low volatility phases. This analysis can be seen in chapter \ref{Extrem_Val}, which compares all used models.
 
If we now look at the signal plot, we see that the predicted signals often fluctuate, this means that the model is very sensitive to the high volatility. Occasionally, longer phases are also visible which the model remains long. 

\newpage

\begin{figure}[ht]
\begin{centering}
\includegraphics[width=.8\linewidth]{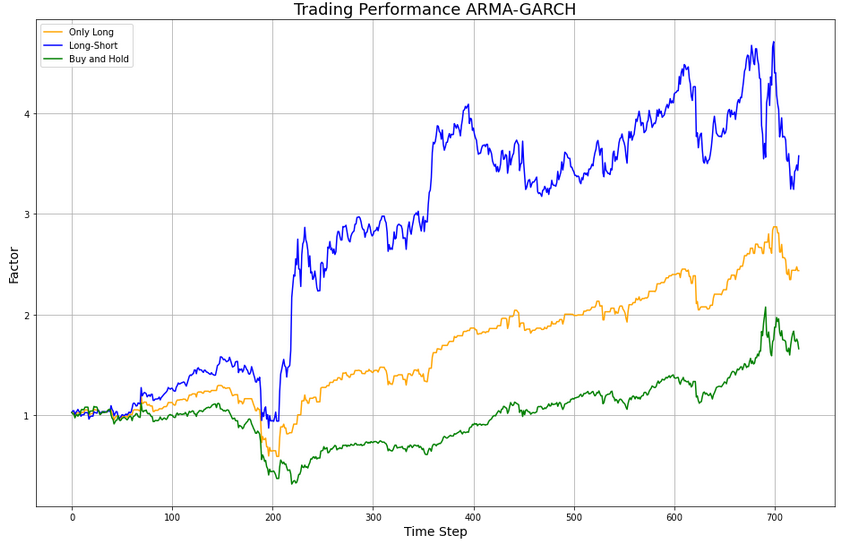}
\caption{ARCH-GARCH Performance-Plot}
\label{Fig9}
\end{centering}
\end{figure}

Figure \ref{Fig9} shows the performance plot of the three strategies (Only Long, Long-Short and Buy and Hold). Over a period of approximately 2 years, the ARMA-GARCH strategy has performed significantly better than the Buy and Hold strategy. Furthermore, in the case of the Long-Short strategy, it can be seen that it has produced the largest gain between 2020-02 and 2020-06. This shows that the ARMA-GARCH model is good at capturing high volatility phases in a turbulent market such as the Covid 19 crisis. 

The long-short strategy performed best with a Sharpe ratio of 0.88 and a profit factor of 3.82. The strategy significantly outperformed the buy and hold, which achieved a Sharpe Ratio of 0.36 and a Profit Factor of 1.73. The Only Long strategy with a Sharpe Ratio of 0.77 and a Profit Factor of 2.55 has also significantly outperformed the Buy and Hold strategy. Thus, one last comment to make about this model is that the ARMA-GARCH model performed better in choppy markets than in calm ones.

\begin{table}
\hfill
\begin{centering}
\caption{Performance of ARMA(1,1)-GARCH(1,1)}
\begin{tabular}{cccc}
\toprule[0.1pt]
\textbf{Trading strategy} & \textbf{Sharpe Ratio} & \textbf{Profit Factor}\\\addlinespace
\midrule[0.1pt]\addlinespace
\textbf{Buy and Hold}   & 0.36 & 1.73\\\addlinespace
\textbf{Only Long} & 0.77 & 2.55 \\\addlinespace
\textbf{Long Short} & 0.88 & 3.82 \\\addlinespace
\bottomrule[0.1pt]\addlinespace[2pt]
\label{Tab4}
\end{tabular}\par
\end{centering}
\end{table}

\newpage

\subsubsection{Cross Signal}

Cross signal trading is a very common method used by chart traders. In this example, a very simple model was developed. Using the close price, two simple moving averages (SMA) were formed. The first with a 3 week interval (15 days) and the second with a 9 week interval (60 days). SMA are so called smoother's which makes sure that the time series are not so volatile. The larger the SMA the smoother the time series. Thus, it is possible that strong trends become apparent. The problem with large SMA is that the time series takes a long time to recognize a change in direction. Therefore it is important to work with two such SMA. After creating two SMA's, the trading rules have to be determined. In this example very simple rules were determined. 

The trading rules are:
If the MA15 crosses the MA60 from the bottom to the top, this is a buy signal. However, if the MA15 crosses the MA60 from the top to the bottom, it is a sell signal. This method belongs to the trend strategies. In the first model only buy or sell was enabled, in the second model also the shortselling.

\begin{figure}[ht]
\begin{centering}
\includegraphics[width=.8\linewidth]{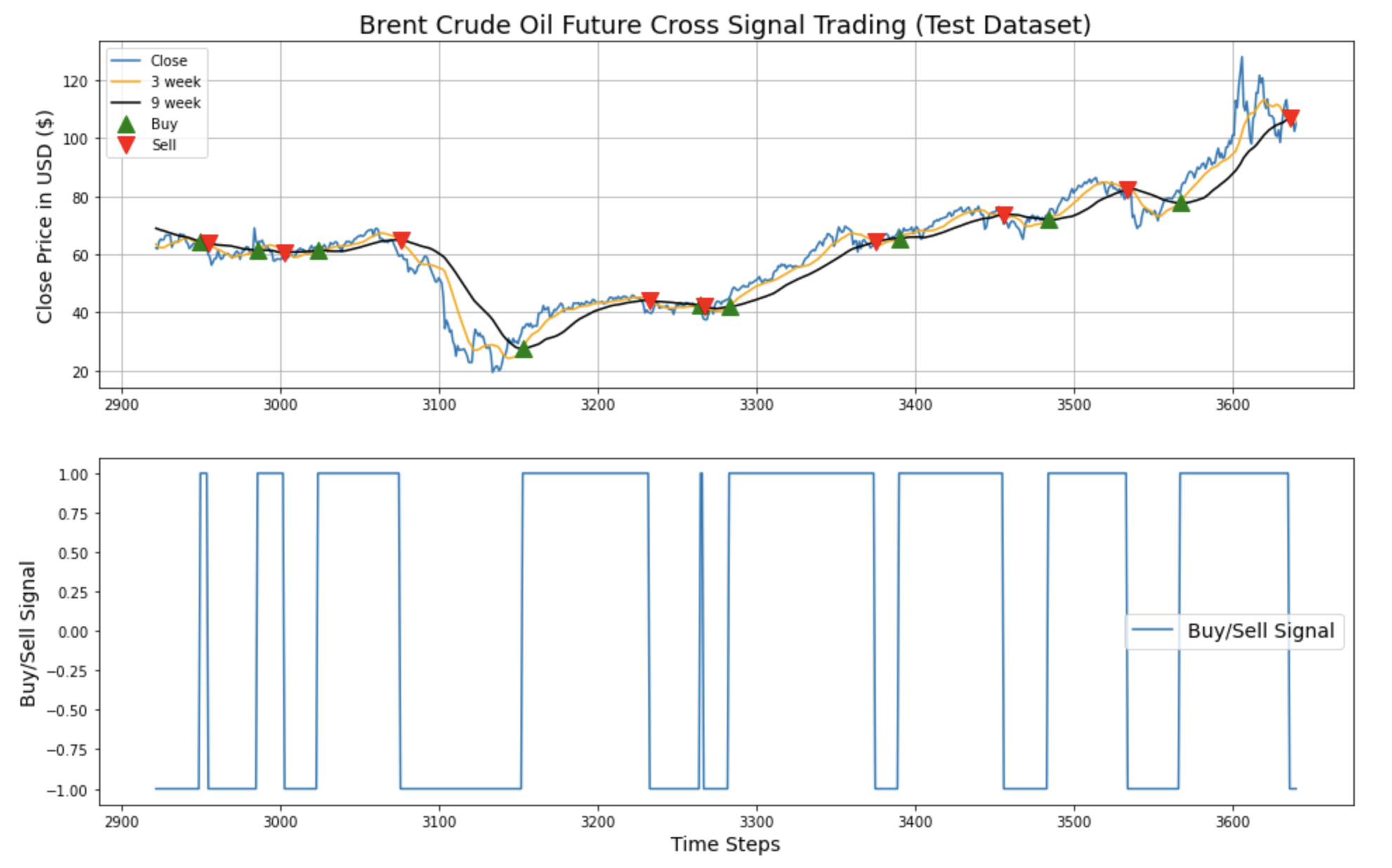}
\caption{Cross Signal Trading with Testdata}
\label{Fig10}
\end{centering}
\end{figure}

In figure \ref{Fig10} you can see the original time series in blue and the MA15 and MA60 in orange and black. The green arrows signal a buy signal and the red a sell signal. In the first phase of the original time series there is no clear trend, here we have an up and down movement of the time series with an average of 60 USD. In such phases, cross signal trading is not very performant because this trading strategy needs trends to achieve good results. After this first phase, the time series starts to have clear trends and here the cross signal trading can make good decisions. 

\newpage

\begin{figure}[ht]
\begin{centering}
\includegraphics[width=.8\linewidth]{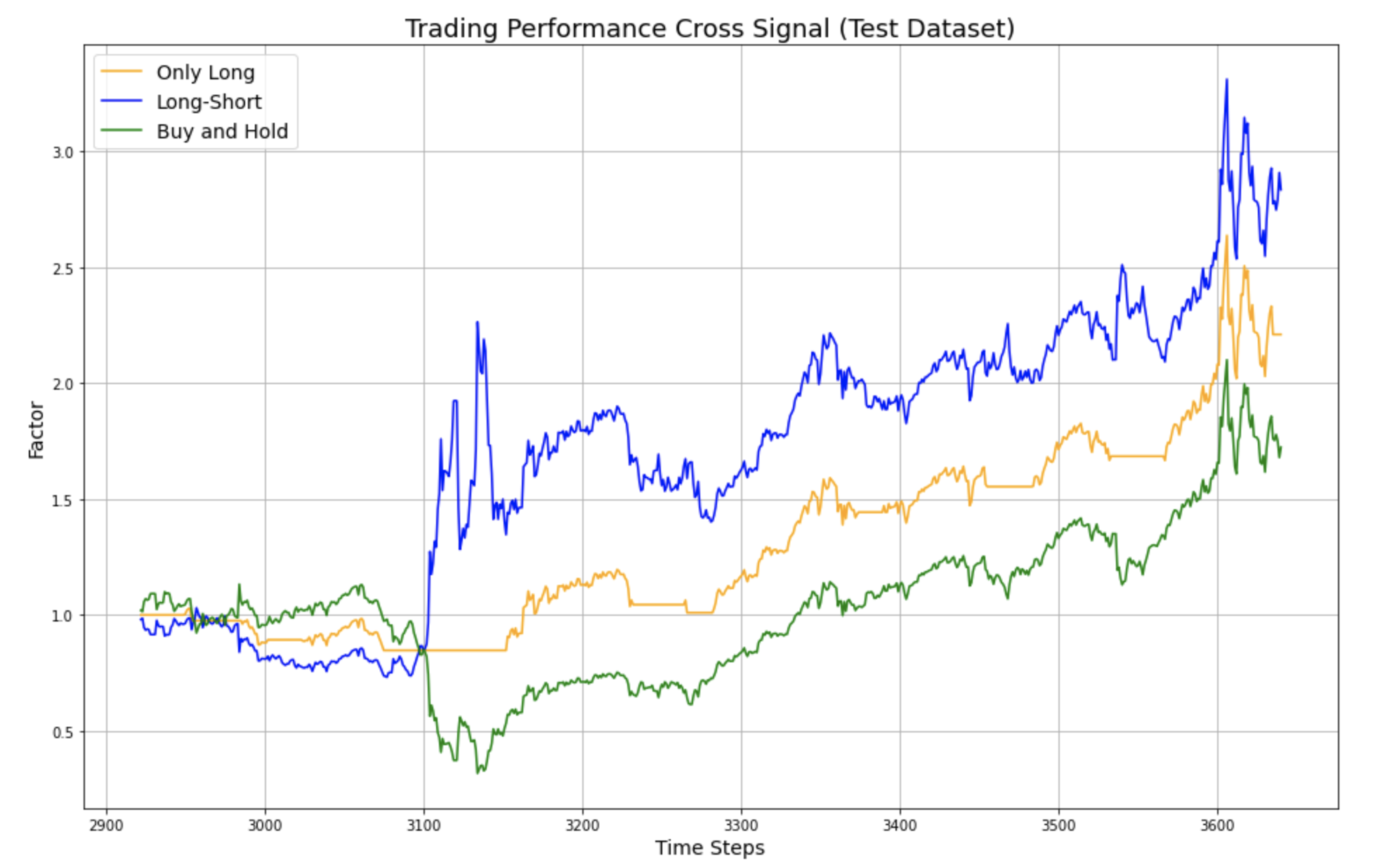}
\caption{Cross Signal Trading Performance Plot}
\label{Fig11}
\end{centering}
\end{figure}

Also in the performance plot (figure \ref{Fig11}) the initial phase is well visible. Only-Long as well as Long Short perform poorly. After that, a clear upward movement in the performance is evident and here you can also generate good returns. If you look closely, you can see that the only-long strategy in certain phases completely exits. This can have a positive or negative impact on performance. Nevertheless, it is a good strategy because it has a lower risk than the long short strategy. 

Both strategies perform better than the buy and hold which can be seen in table \ref{Tab5}. Only Long achieves a better Sharp Ratio than Long Short which confirms the statement that the Only Long strategy has a lower risk than Long Short. The profit factor is highest with the Long Short strategy, this is because this strategy is also profitable when the original time series has a downward movement. 
  
\begin{table}
\hfill
\begin{centering}
\caption{Perforance Table of Cross Signal trading}
\begin{tabular}{cccc}
\toprule[0.1pt]
\textbf{Trading strategy} & \textbf{Sharpe Ratio} & \textbf{Profit Factor}\\\addlinespace
\midrule[0.1pt]\addlinespace
\textbf{Buy and Hold}   & 0.36 & 1.73 \\\addlinespace
\textbf{Only Long} & 0.95 & 2.21 \\\addlinespace
\textbf{Long Short} & 0.69 & 2.83 \\\addlinespace
\bottomrule[0.1pt]\addlinespace[2pt]
\label{Tab5}
\end{tabular}\par
\end{centering}
\end{table}

\newpage

\subsection{Machine Learning Models}

This chapter contains the results of the machine learning models LSTM, Random Forest, Support Vector Machine Regression (SVM) and k-Nearest-Neighbor (kNN). First, all predictions of the respective models are presented, followed by a classification report, which evaluates the accuracy of the predictions. Then the Feature Importance is performed, which shows how much each feature contributes to the model. Then the performance of the individual models is presented and finally the cross validation and a extreme value analysis were performed. 

\subsubsection{The structure of the models and their predictions}

In the first step, we have to determine the model structure and their predictions. This will be shown in this section. 

\textbf{Long short-term memory}\label{Chapter_LSTM}

The LSTM networks can solve the problem of vanishing gradient, which ordinary RNN networks have. This thanks to the property that these networks can track the long-term dependencies in data. For more detailed examples see "Deep Learning; Adaptive Computation and Machine Learning Series" which was published by Goodfellow et al. 2016 \cite{goodfellow2016deep}. We first fit a GARCH(1,1) model to find the number of lags using the standardized residuals. In the previous chapters, the distribution of Brent crude oil prices was shown to follow a Students-t distribution, so the GARCH model was fitted with this assumption. The standardized residuals should not show any dependence structure, if this should be the case, then we know how many lags are needed to explain $t+1$. The ACF analysis showed that the data must be lagged for 39 days. For the LSTM algorithm we create lagged data-sets so that we have a kind of a linear function who train the neural network. The LSTM creates the optimal parameters in each node to predict the next day with the past data. After we have to split the Close Price in a train and test data-set with a 80:20 ratio. The data used has been normalized. This is to improve the performance of the model, bringing the values of numerical columns to a common scale. Hence, the training data set was transformed to a scale of zero and one. The architecture of the LSTM model consists of two LSTM layers and two Dense layers. The LSTM layers each have 128/64 neurons and the Dense layers each have 25/1 neuron, the latter being the output neuron. Each of these LSTM layers uses the activation function tanh and the recurrent activation function sigmoid. The tanh function allows the negative values to be retained. Additionally, "return\_sequences=True" was set in the programming, this is required for stacking LSTM layers so that the subsequent LSTM layer has a three dimensional sequence input. The optimization of the model was optimized with the Adam optimizer (Kingma and Ba, 2017) \cite{kingma2014adam} , which is a stochastic gradient descent optimizer with momentum and the loss was set as mean square error. Then the model is trained 5 epochs, the epochs describe the number of times the learning algorithm is run through the entire training set. The batch size is set to one so that the optimizer can work more efficiently. The hyperparameters are listed in the following table \ref{Tab6}:

\begin{table}
\hfill
\begin{centering}
\caption{Values of hyperparameter}
\begin{tabular}{cccc}
\toprule[0.1pt]
\textbf{Hyperparameter} & \textbf{Selected Value}\\\addlinespace
\midrule[0.1pt]\addlinespace
No. hidden layers   & 4 \\\addlinespace
No. neurons & 128 / 64 / 25 / 1 \\\addlinespace
Activation function & tanh \\\addlinespace
Recurrent activation function & sigmoid \\\addlinespace
Optimizer & Adam \\\addlinespace
Dropout rate  & 0 \\\addlinespace
Batch size & 1 \\\addlinespace
Loss function & MSE \\\addlinespace
Learning rate & 0.001 \\\addlinespace
\bottomrule[0.1pt]\addlinespace[2pt]
\label{Tab6}
\end{tabular}\par
\end{centering}
\end{table}

\begin{figure}[ht]
\begin{centering}
\includegraphics[width=.65\linewidth]{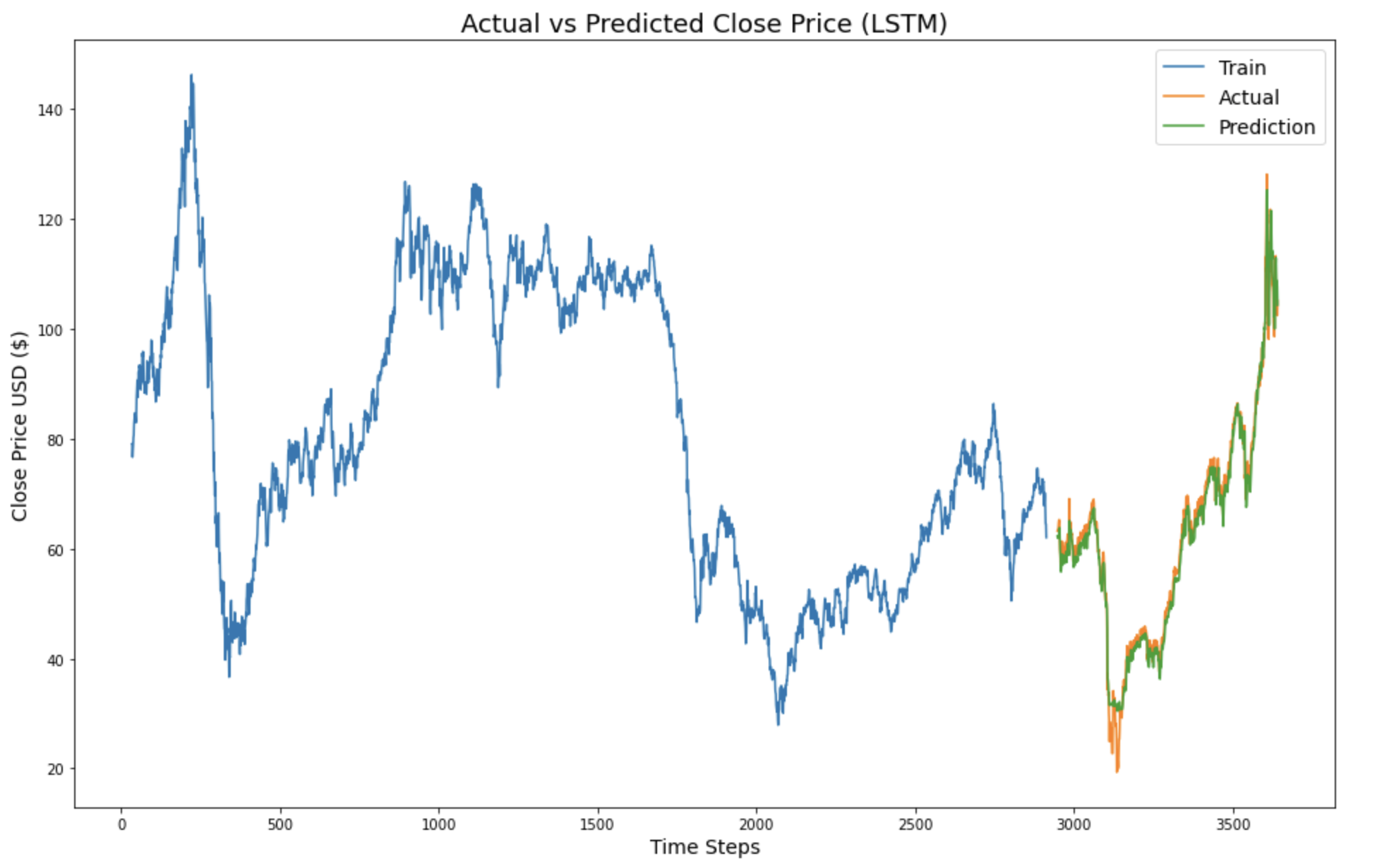} 
\includegraphics[width=.65\linewidth]{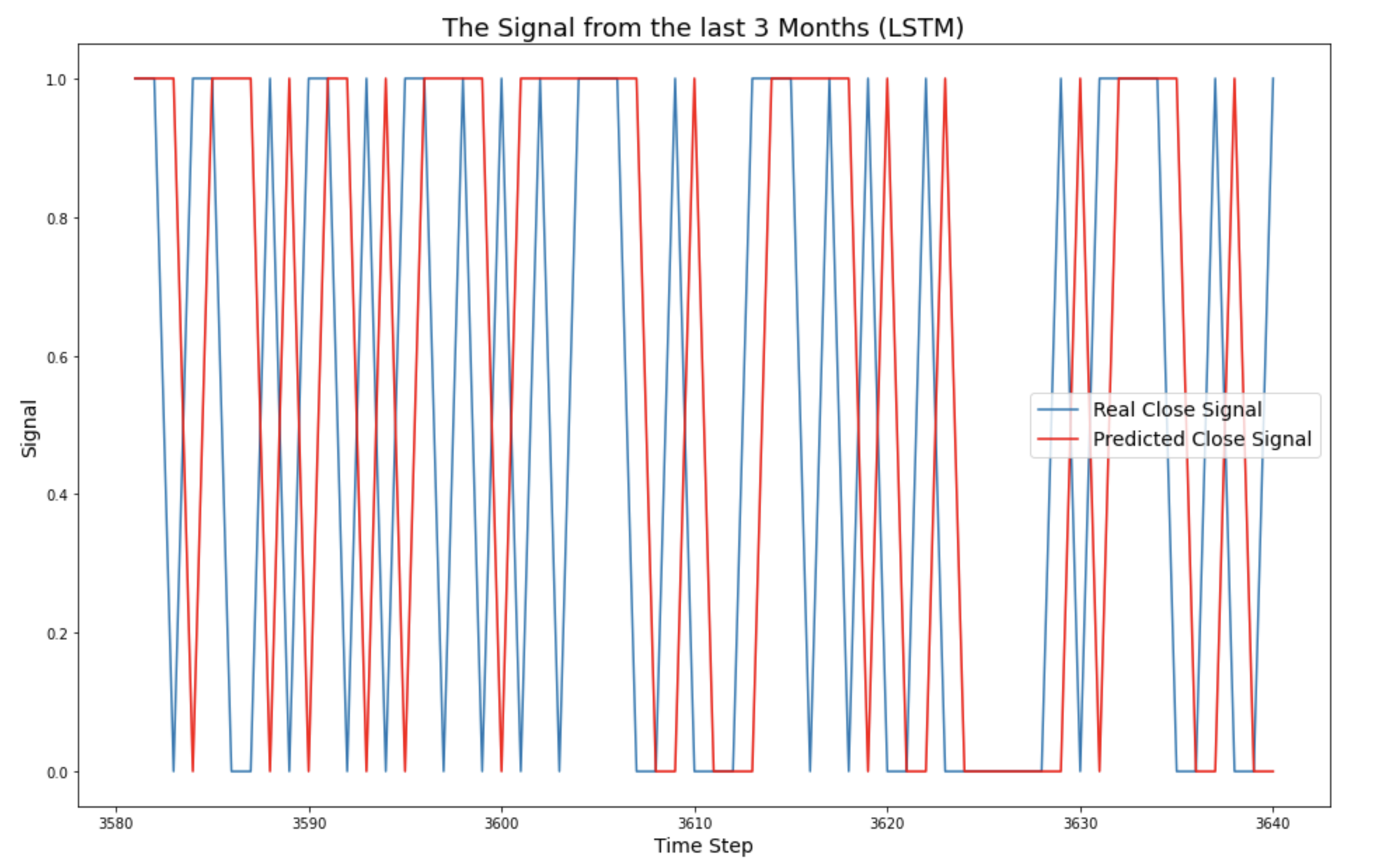} 
\caption{LSTM Prediction and LSTM Signals}
\label{Fig12}
\end{centering}
\end{figure}

The predictions of the LSTM model can be seen in figure \ref{Fig12}. To get the predicted signals, we had to convert the predicted closing price into log-returns. Afterwards these log-returns were converted into 0 and 1 signals with the signum operator.  

The model correctly predicted the signals with a accuracy of 51.81\%. Furthermore, it can be seen that the model tends to be positioned long in calm market phases and sings between long and short in volatile phases. 

\newpage

\textbf{Random Forest}

This section presents the results of predicting trading signals for Brent crude oil. The Random Forest classifier works very simply: it creates a set of decision trees from a randomly selected subset of the training set. Then, the votes of the different decision trees are combined to determine the final class of the test object. In this work, a Random Forest was used for classification. Financial indicators were defined as features to capture the characteristics of Brent crude oil, while the Random Forest (RF) model is used to train on the training data-sets according to a certain classification criterion and make predictions for the next trading day.\cite{luber_litzel_2020} 

For the Random Forest model, the data set was split into training and test data in the ratio of 80/20. Then the tuning of the hyperparameters was performed in Python using the RandomizedSearchCV command. The following table \ref{Tab7} shows what the command outputted.

\begin{table}
\hfill
\begin{centering}
\caption{Values of hyperparameter (Random Forest)}
\begin{tabular}{cccc}
\toprule[0.1pt]
\textbf{Hyperparameter} & \textbf{Selected Value}\\\addlinespace
\midrule[0.1pt]\addlinespace
No. of trees   & 169 \\\addlinespace
No. of features to consider at every split & 2 \\\addlinespace
Maximum number of levels in tree & 4 \\\addlinespace
Minimum number of samples required to split a node & 49 \\\addlinespace
Minimum number of samples required at each leaf node & 1 \\\addlinespace
Criterion  & entropy \\\addlinespace
\bottomrule[0.1pt]\addlinespace[2pt]
\label{Tab7}
\end{tabular}\par
\end{centering}
\end{table}

\begin{figure}[ht]
\includegraphics[width=.50\linewidth]{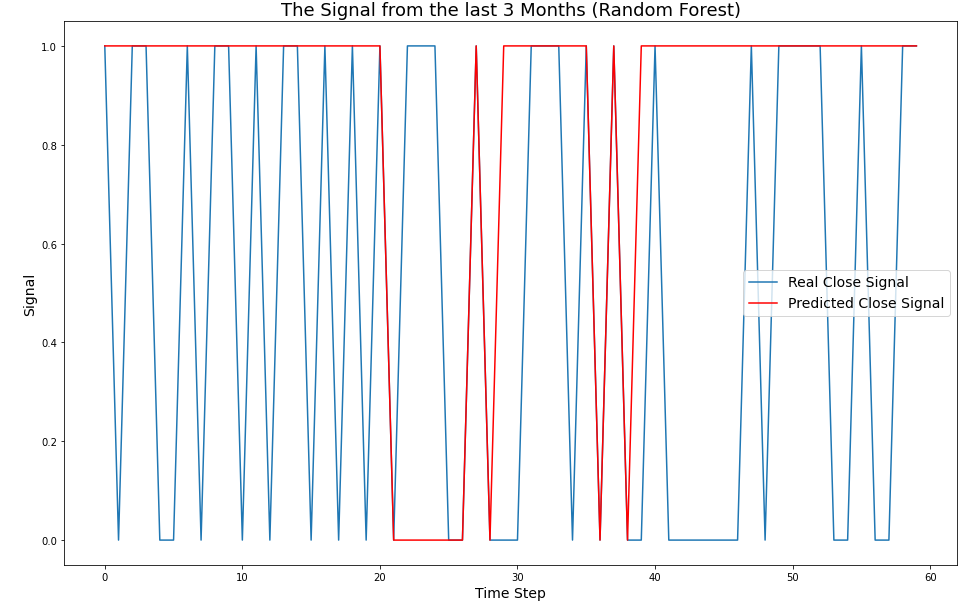}
\hfill
\includegraphics[width=.45\linewidth]{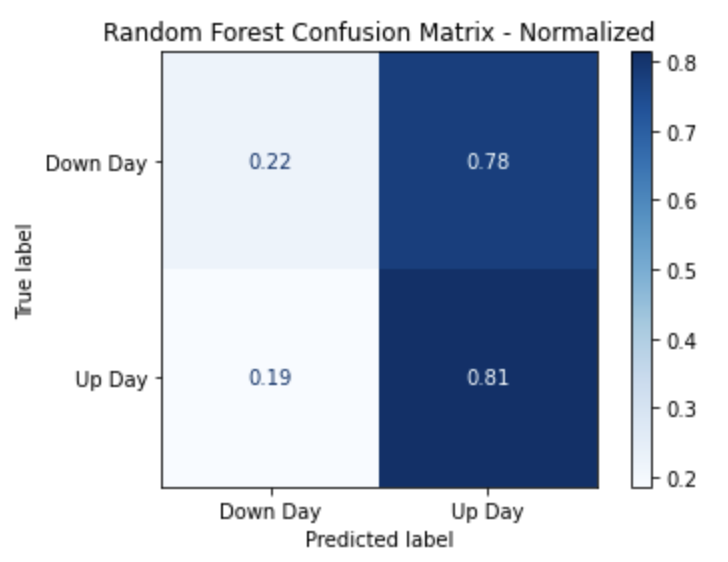}
\caption{Signal Plot (left) and Confusion Matrix (right)}
\label{Fig13}
\end{figure}

The Random Forest model was initialized with these parameters and then trained. Then a prediction of the labels was made with the test data. In figure \ref{Fig13} you can see the signal predicted by the random forest model and the confusion matrix. It is obvious that the model predicts up days in the largest part and only a few down days. This also explains the high recall value for the Up Days, which can be seen in a later chapter.The Random Forest model achieves the highest accuracy with 54.73\%.  

\newpage

\textbf{Support Vector Machine}

For signal prediction, a SVM regression model with a radial basis kernel was used. The SVM can solve the same problems as ordinary linear regression. The main difference is that SVM can also solve non-linear problems. By tuning hyperparameters, it is possible to develop very flexible and robust models. In the SVM algorithm, there are two specific parameters that must be estimated. These are $C$ and $\epsilon$. The parameter $C$ is a regularization parameter, which regulates the distance $\xi$ to the points outside the hyperplane. $\epsilon$ determines the width of the hyperplane. To find the best value for $C$, a list of 100 values from 1 to 100 was generated in Python and for $\epsilon$ a list of 0.1 to 50.1 with a sequence of 0.1. Subsequently, the best values for $C$ and $\epsilon$ are searched with the hyperparameter Tuning. For this model, the features from chapter 4 were used. The main difference to the other models is that the label is now based on the $Closing Price_{t-1}$. 
For this model the data was again splitted into training and test data with a 80:20 ratio. Afterwards a hyperparameter tuning with cross validation was done, this was done in Python with the command RandomizedSearchCV. The following table \ref{Tab8} shows the best parameters that lead to the best model.

\begin{table}
\hfill
\begin{centering}
\caption{Values of hyperparameter (SVR)}
\begin{tabular}{cccc}
\toprule[0.1pt]
\textbf{Hyperparameter} & \textbf{Selected Value}\\\addlinespace
\midrule[0.1pt]\addlinespace
Shrinking   & True \\\addlinespace
Gamma & Auto \\\addlinespace
$\epsilon$ & 22.4 \\\addlinespace
C & 19 \\\addlinespace
Cache size & 41 \\\addlinespace
\bottomrule[0.1pt]\addlinespace[2pt]
\label{Tab8}
\end{tabular}\par
\end{centering}
\end{table}

\begin{figure}[ht]
\includegraphics[width=.5\linewidth]{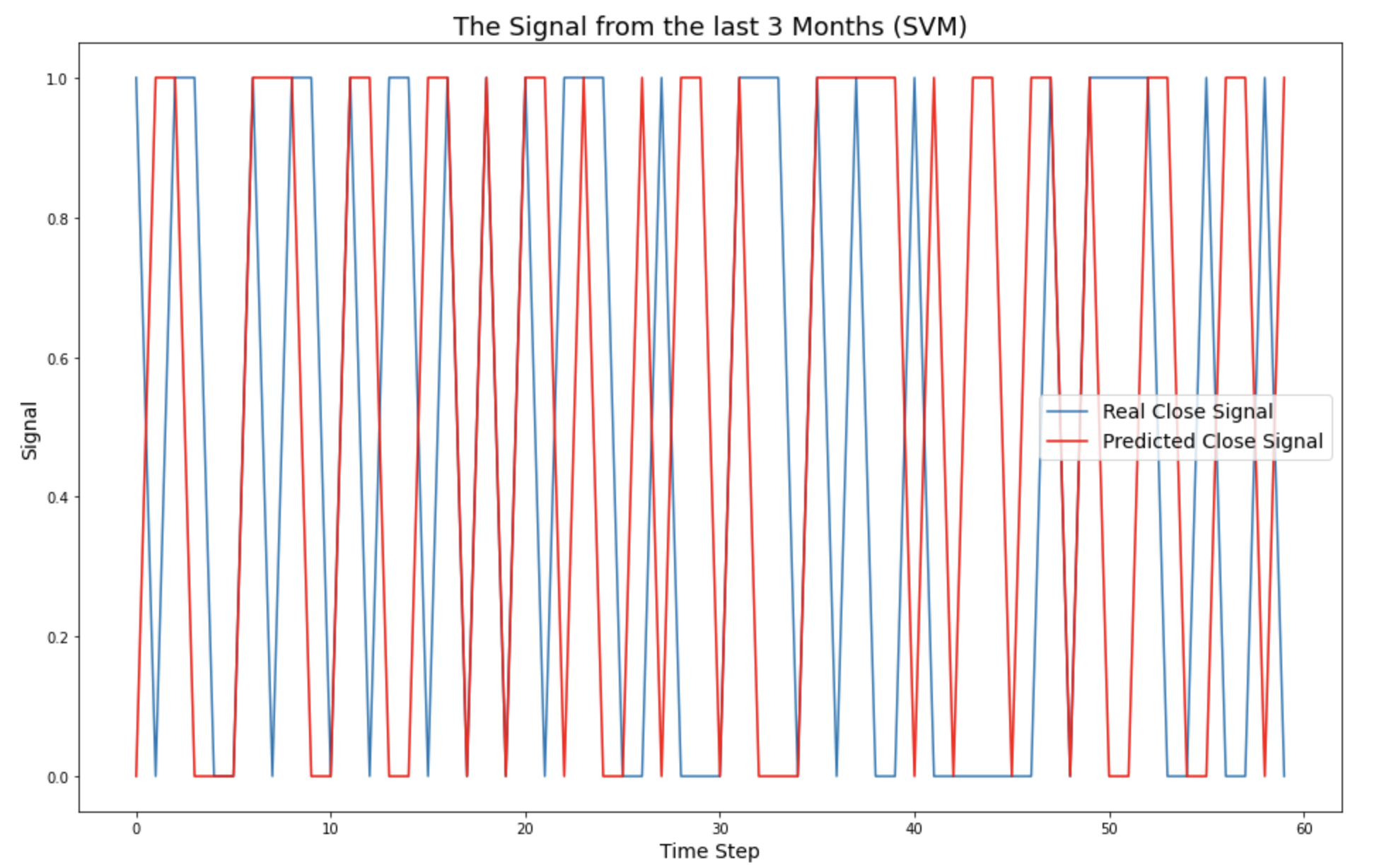}
\hfill
\includegraphics[width=.45\linewidth]{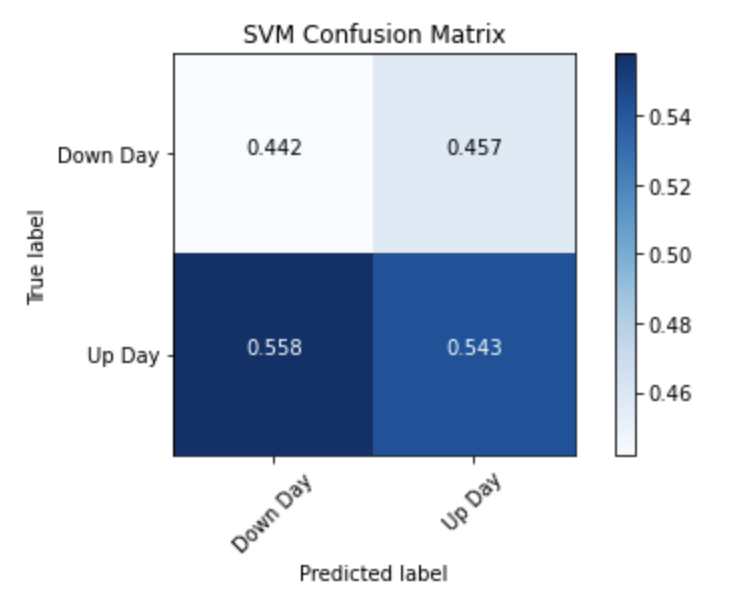}
\caption{Signal Plot (left) and Confusion Matrix (right)}
\label{Fig14}
\end{figure}

With the hyperparameters from the table \ref{Tab8} the SVM algorithm was trained. After this a prediction was made of the labels with the test data. In the figure \ref{Fig14} the predicted signals and the confusion matrix are shown. The SVM model achieves an accuracy of 49.18\%. The SVM model rarely has long periods in which it is in the same class (0 or 1), it fluctuates constantly. This is also shown by the confusion matrix, the two classes are approximately equally distributed, the class Up day (i.e. 1) is predicted a bit more often than the class Down day (i.e. 0). 

\newpage

\textbf{k-Nearest Neighbor}

In this section, the K-Nearest Neighbor (kNN) model is applied to Brent crude oil futures to predict a trading signal for a sample of test data. The kNN was applied as a classification model. All financial indicators presented in chapter \ref{financ_indi} were used as feature inputs. The generated signals from the whole data-set were used as labels. Furthermore, the input features were normalized, these were scaled between 0 and 1, so that the algorithm can work more efficiently and the calculation of the distances is simplified. The Manhattan distance was used, then the data was splitted into test and training data. Hyperparameter tuning table \ref{Tab9} was done to see which hyperparameter leads to the best fit of the model. Then the model was validated and checked how well it performed. This was done using accuracy score, which we rely on it in a later step. To get more details about the performance a classifications report was made. Also the distribution of the predictions is looked at with the confusion matrix. A Feature Importance analysis shows the explanatory power of the input features on the model, this will also be looked at later in chapter \ref{feat_impo}. Finally, the final results are discussed. For the model, the data set was split into training and test data in an 80:20 ratio. Then the hyperparameter tuning was done in Python with the RandomizedSearchCV command. Therefor, lists were created for the number of leafs, the distances, for the weights and for the determination of K, which were then tested to obtain the best possible parameters. The final result is shown in the table below. 

\begin{table}
\hfill
\begin{centering}
\caption{Hyperparameters after tuning the kNN Model}
\begin{tabular}{cc}
\toprule[0.1pt]
\textbf{Tuning Parameter} & \textbf{Selected Value}\\\addlinespace
\midrule[0.1pt]\addlinespace
Leaf size & 15 \\\addlinespace
Distance & Manhattan \\\addlinespace
Weight & Distance \\\addlinespace
No. K & 5 \\\addlinespace
Algorithm & Auto \\\addlinespace
\bottomrule[0.1pt]\addlinespace[2pt]
\label{Tab9}
\end{tabular}\par
\end{centering}
\end{table}

\begin{figure}[ht]
\includegraphics[width=.5\linewidth]{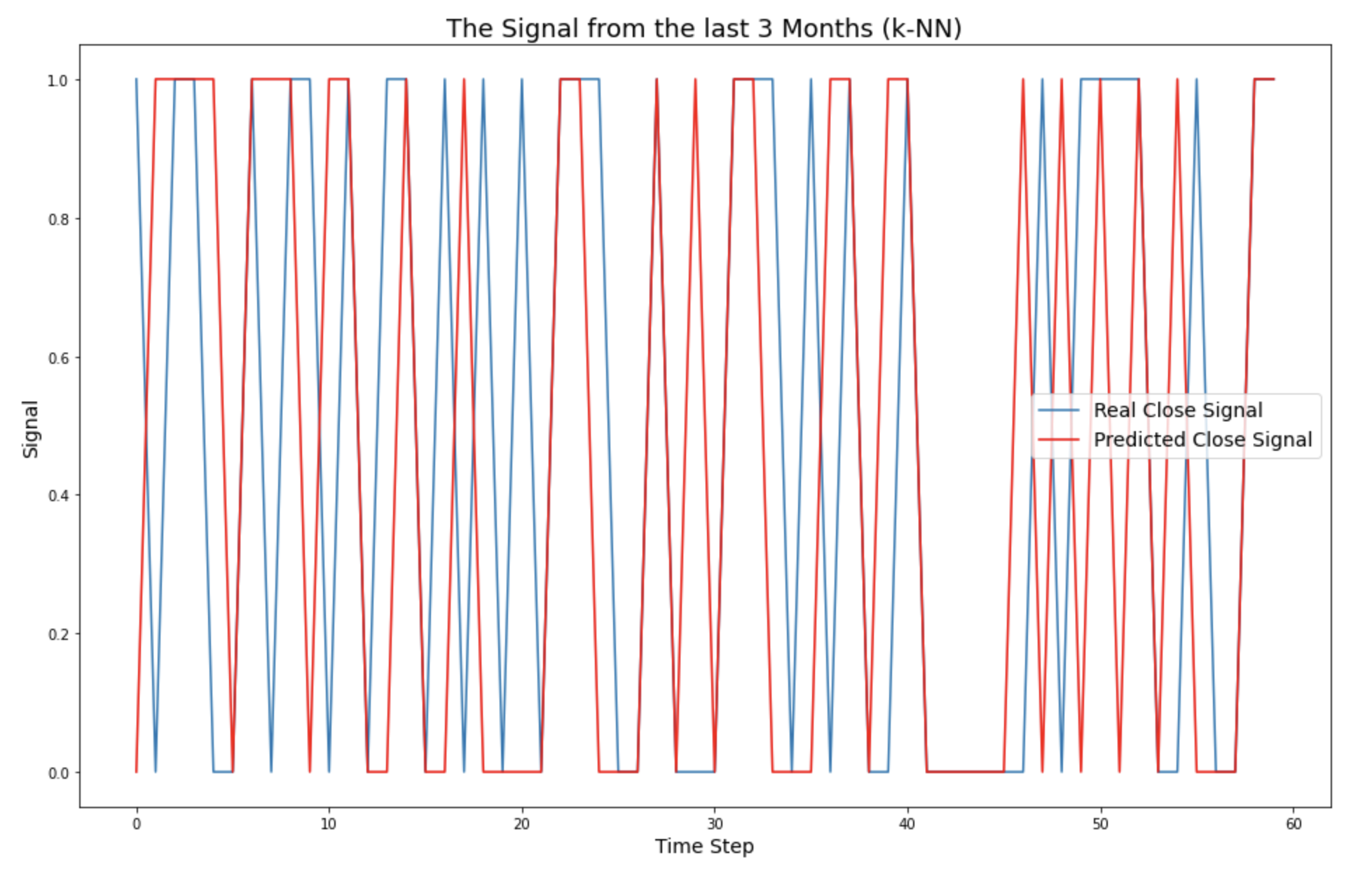}
\hfill
\includegraphics[width=.5\linewidth]{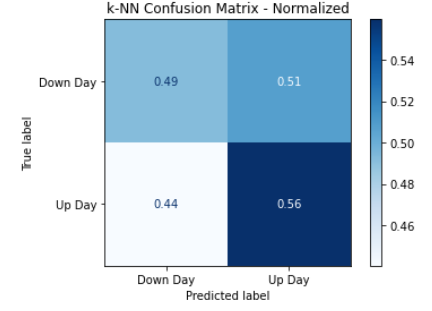}
\caption{Signal Plot (left) and Confusion Matrix (right)}
\label{Fig15}
\end{figure}

In the figure \ref{Fig15} the predicted Signals vs. the real signals and the confusion matrix are shown. The signals fluctuate very firmly with this model. Occasionally, there are periods in which the model remains in the same class for a certain time. However, the confusion matrix shows that the two classes (0 and 1) are very balanced. 

\newpage

\subsubsection{Classification Report}

To measure the quality of predictions of the Machine Learning Models a classification report was made. This is shown in the table \ref{Tab10}. The report shows the main classification metrics accuracy, recall, f1 score and support per class. The metrics are calculated using true and false positives, true and false negatives. The precision tells how many of all positively classified instances were classified correctly. Now consider the recall, this describes what percentage of all cases that were actually positive were correctly classified. For the F1-score, the best value is 1 and the worst 0. How all this was calculated can be seen in the appendix. 

The precision of the Random Forest model is for Down Days slightly better then for Up Days, which doese not mean, that the model can better predict the Down Days, if we rely on the recall, which significantly better performed for UP days than for Down days, which means that the Random Forest model can predict the Up days better than the Down days. If we are looking on the precision of the SVM model we see that the precision is bit better for the Up days than for the Down days, which means that the model predicts the Up days better than the Down days. Because the recall makes no significantly better difference between the Up and Down Days.

If we look at the kNN Model, the class Down Day, which corresponds to class 0, we see that it was classified correctly 47.6\% of the time, compared to Up Day, which corresponds to class 1. The precision was classified 57.5\% correctly. This means that if the model predicts class 1, we can be more confident that it has been correctly predicted. Now consider the recall, for class 0 the recall was 49.2\% and for class 1 56\%, this means that for class 1 56\% of the positive cases were actually positive. When looking on the f1-score th class 1 is better predicted than class 0. 

The conclusion of this classification report is that if the model predicts class 1, we can trust it more than if it predicts class 0. 

\begin{table}
\hfill
\begin{centering}
\caption{Classification Report of all ML models}
\begin{tabular}{cccccr}
\toprule[0.1pt]
               &     \textbf{Precision} & \textbf{Recall} & \textbf{F1 score} &  \textbf{Support}\\\addlinespace
\midrule
\emph{\textbf{Down Day}}&         &         &         &         &         \\
\addlinespace
\hspace{0.25cm} LSTM & 0.460 & 0.389 & 0.422 & 311 \\
\addlinespace
\hspace{0.25cm} RF &  \textbf{0.490}  &   0.223 & 0.307  &  337 \\
\addlinespace
\hspace{0.25cm} SVM & 0.441 & \textbf{0.492} &   0.465 & 327 \\
\addlinespace
\hspace{0.25cm} kNN & 0.476 & \textbf{0.492} & \textbf{0.484} & 327 \\
\addlinespace
\vspace{0.1em} \\ \emph{\textbf{Up Day}}&         &         &         &                 \\
\addlinespace
\hspace{0.25cm} LSTM & 0.554 & 0.624 & 0.587 & 378 \\
\addlinespace
\hspace{0.25cm} RF &  0.562 &   \textbf{0.811} & \textbf{0.664}  &  402 \\
\addlinespace
\hspace{0.25cm} SVM & 0.543 & 0.491 &   0.516 & 401 \\
\addlinespace
\hspace{0.25cm} kNN & \textbf{0.575} & 0.560 & 0.567 & 402\\
\addlinespace

\bottomrule[0.1pt]\addlinespace[2pt]
\label{Tab10}
\end{tabular}\par
\end{centering}
\end{table}

\newpage

\subsubsection{Feature Importance}\label{feat_impo}

Feature Importance was determined using permutation. The permutation works as follows. The model is run without modification to determine the accuracy. In a second step, the values of one feature are changed at a time to find out how sensitive the model reacts to changes in a specific feature. In this way it can be determined which feature has a large influence on the accuracy.

\begin{figure}[ht]
\begin{centering}
\includegraphics[width=.38\linewidth]{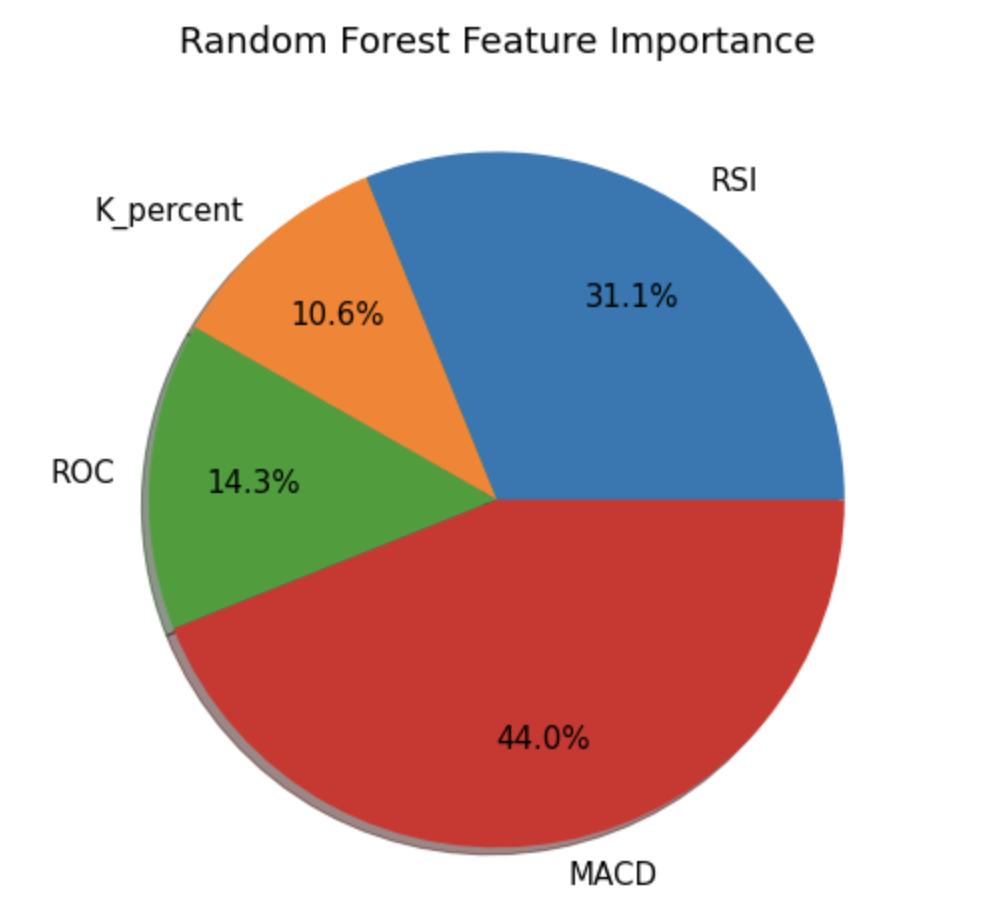}
\includegraphics[width=.38\linewidth]{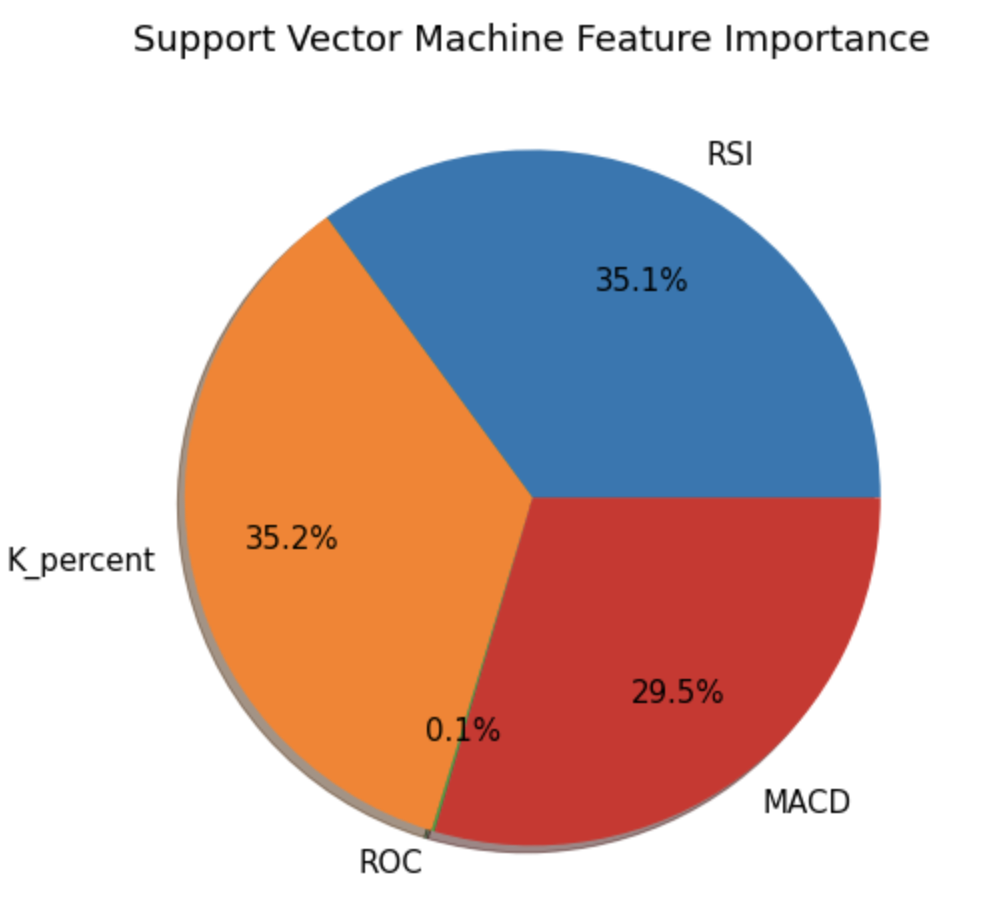}
\includegraphics[width=.38\linewidth]{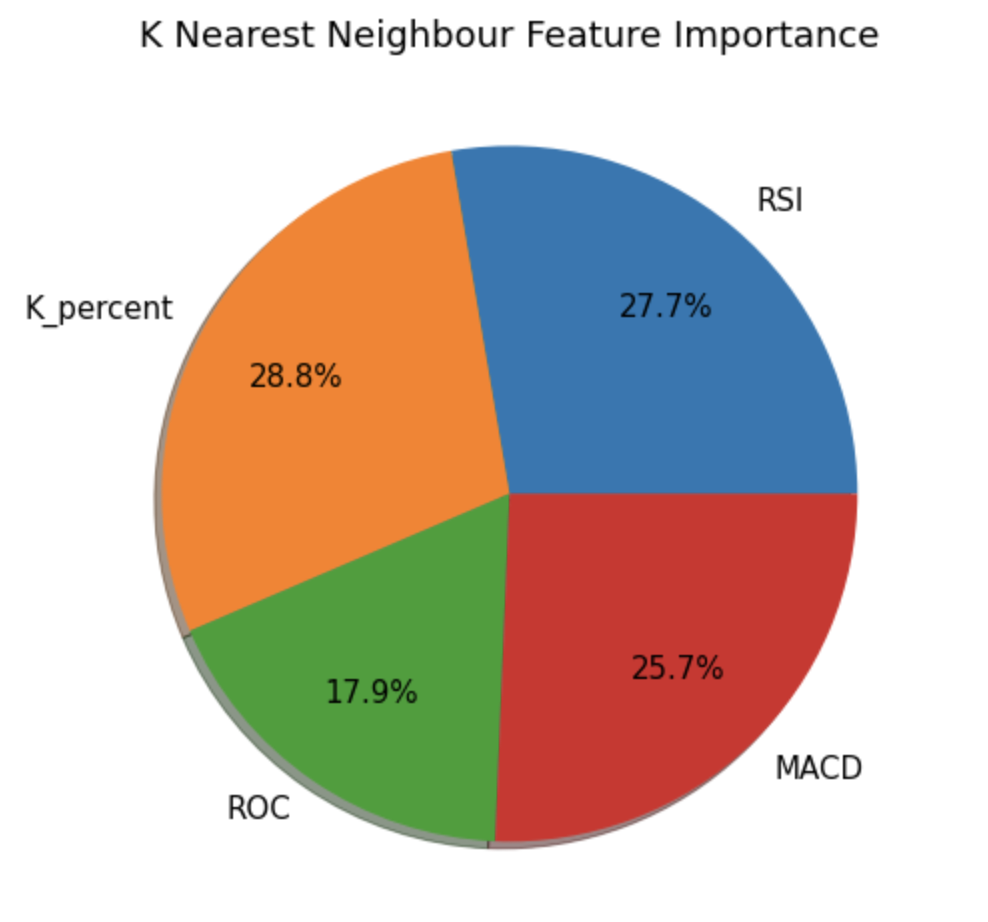}
\caption{Feature Importance}
\label{Fig16}
\end{centering}
\end{figure}

In figure \ref{Fig16} you can see the result of the feature importance of the RF, kNN and SVM model. If we are looking on the RF Plot, all features have an influence on the accuracy with the feature MACD having the largest influence with 44\% and K-Percent with 10.6\% the smallest. It is difficult to say why MACD has the biggest influence on the accuracy. When looking at the distribution of the MACD feature (Section descriptive Statistics) we can see that the distribution is normally distributed with a large tail in the negative direction. Looking at the feature K-Percent it is obvious that the distribution is bimodal. This could also explain why this feature performs worst here. If we have a look on feature importance plot of the SVM model. RSI has the an influence on the model with 35.1\%, followed by K-percent, which has the biggest influence on the model with 35.1\%. MACD also has an influence on the model with 29.5\%. ROC has the smallest influence with 0.1\%. Last, the feature importance plot of kNN is considered. K-percent contributes the most to the model with 28.8\%, followed by RSI with 27.7\%. MACD has 25.7\% share and ROC  with a share of 17.9\%. kNN has the most balanced feature distribution. These proportions are similar for the kNN Model. 

\newpage

\subsubsection{Trading Performance}

\begin{figure}[ht]
\begin{centering}
\includegraphics[width=.7\linewidth]{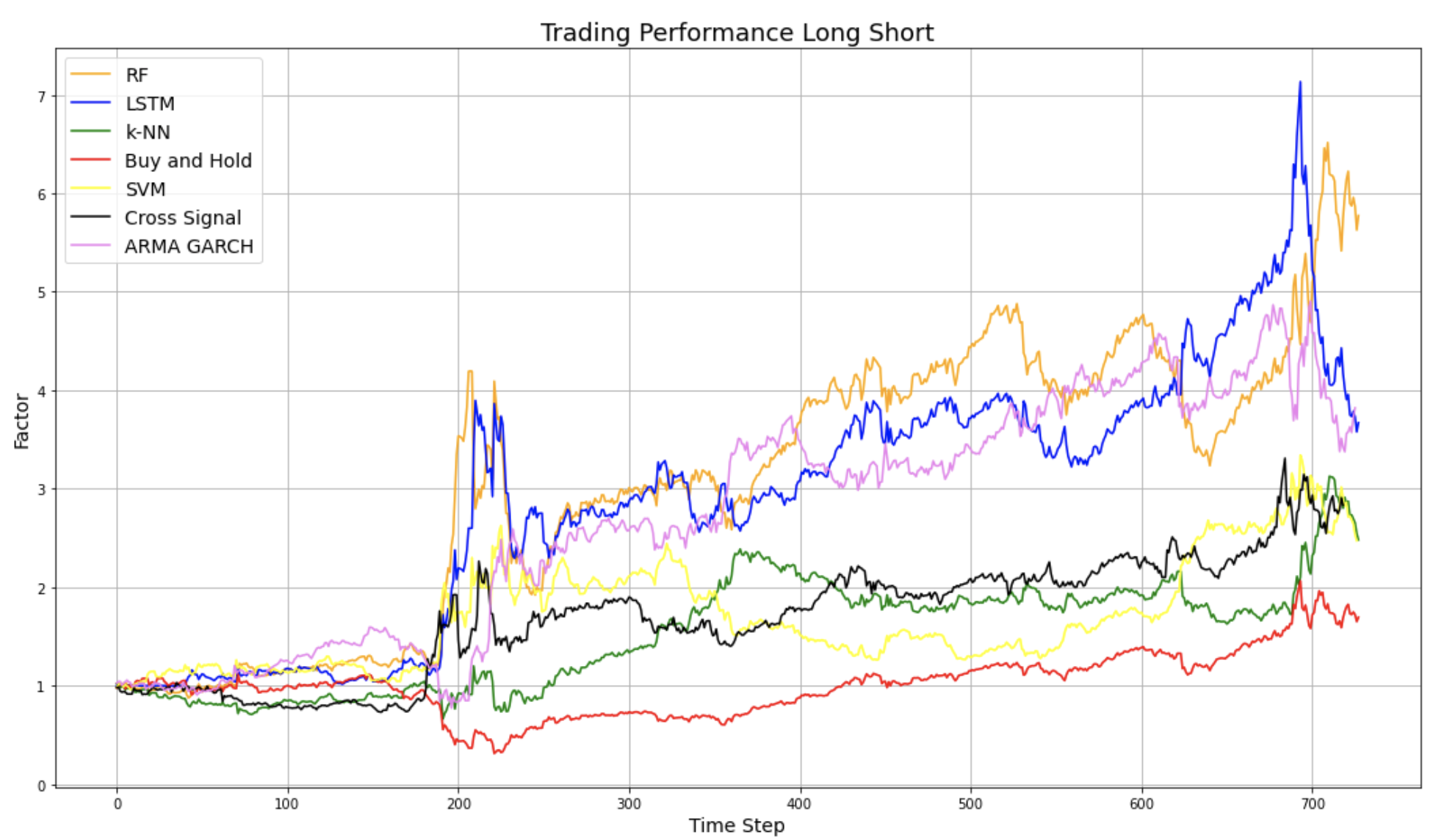} 
\includegraphics[width=.7\linewidth]{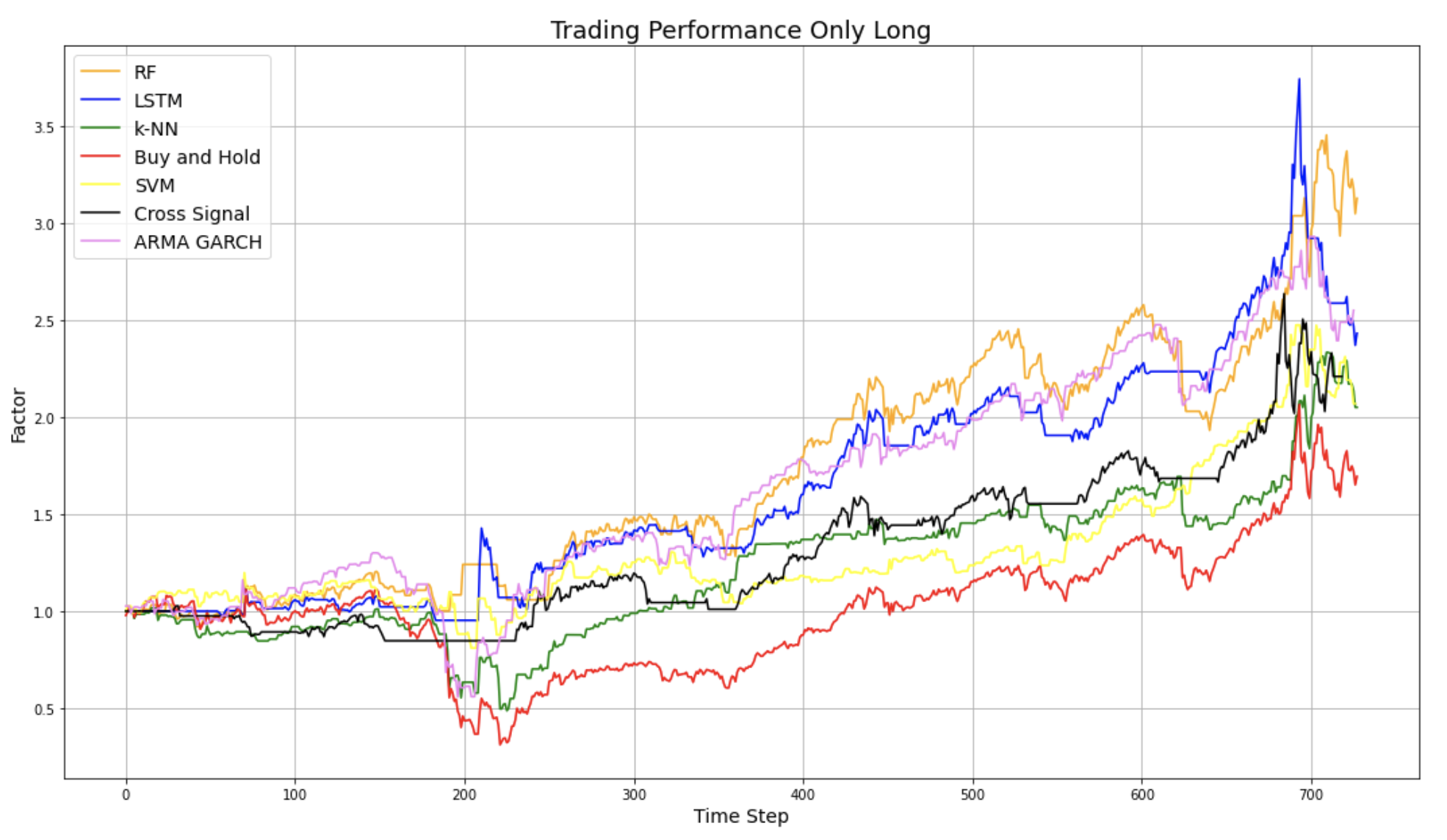} 
\caption{Performance Plot}
\label{Fig17}
\end{centering}
\end{figure}

In this chapter, all performance results, including the benchmark models, are visualized and discussed. In addition, the monthly returns from the test phase are visualized to see in which market phases the different models performed best. In this analysis the profit factor is considered, but the risk of each strategy and performance evaluation is essential. This will be discussed in later chapters, but for now, only the Profit Factor will be discussed.  The first thing to see is that the ARMA-GARCH model performed very well with both the long-short and long-only strategies. This gives us to understand that the ordinary statistical models are still competitive. Again, it gives us to understand that the ML methods still need to be further developed. Nevertheless, we consider the figure \ref{Fig17} with the results. Let's start with the long short strategy. It is pleasing to see that every model outperformed the Buy and Hold strategy. The model that performed the best was the RF model. The profit factor gap with the next best model, namely the ARCH-GARCH, is clearly visible. Surprisingly, the kNN model did not perform as well as hoped. Furthermore, the LSTM model rash at the beginning of the Corona crisis is visible. This shows that the model has recognized the characteristics of the time series at this time very well. However, this positive trend suddenly collapses at the beginning of the Ukraine war and loses almost 50\% of the returns.  Although the Brend Crude price had a positive trend at most of the time, except for an outlier at the very end of the time series, the LSTM model could not capture and detect this. Looking at the Only Long strategy, we see that the performance has a much more constant positive trend and there are very rare performance dips. Again, the RF model has performed the best. And each model has outperformed the buy and hold. Consequently, the risk assessment of the Only Long strategy will be better than that of the Long Short strategy, which intuitively makes sense.  
 
\begin{figure}[ht]
\begin{centering}
\includegraphics[width=.45\linewidth]{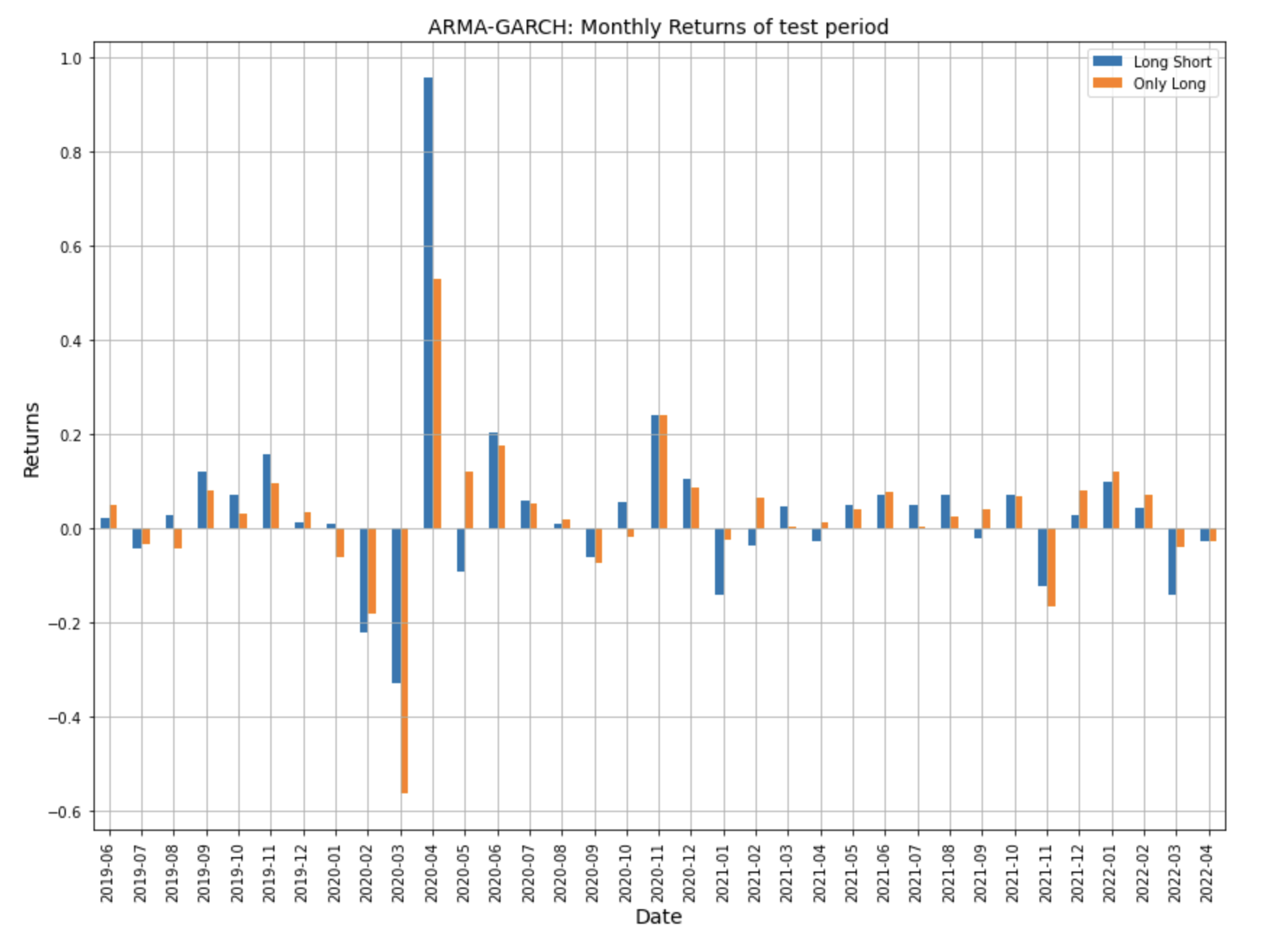}
\includegraphics[width=.45\linewidth]{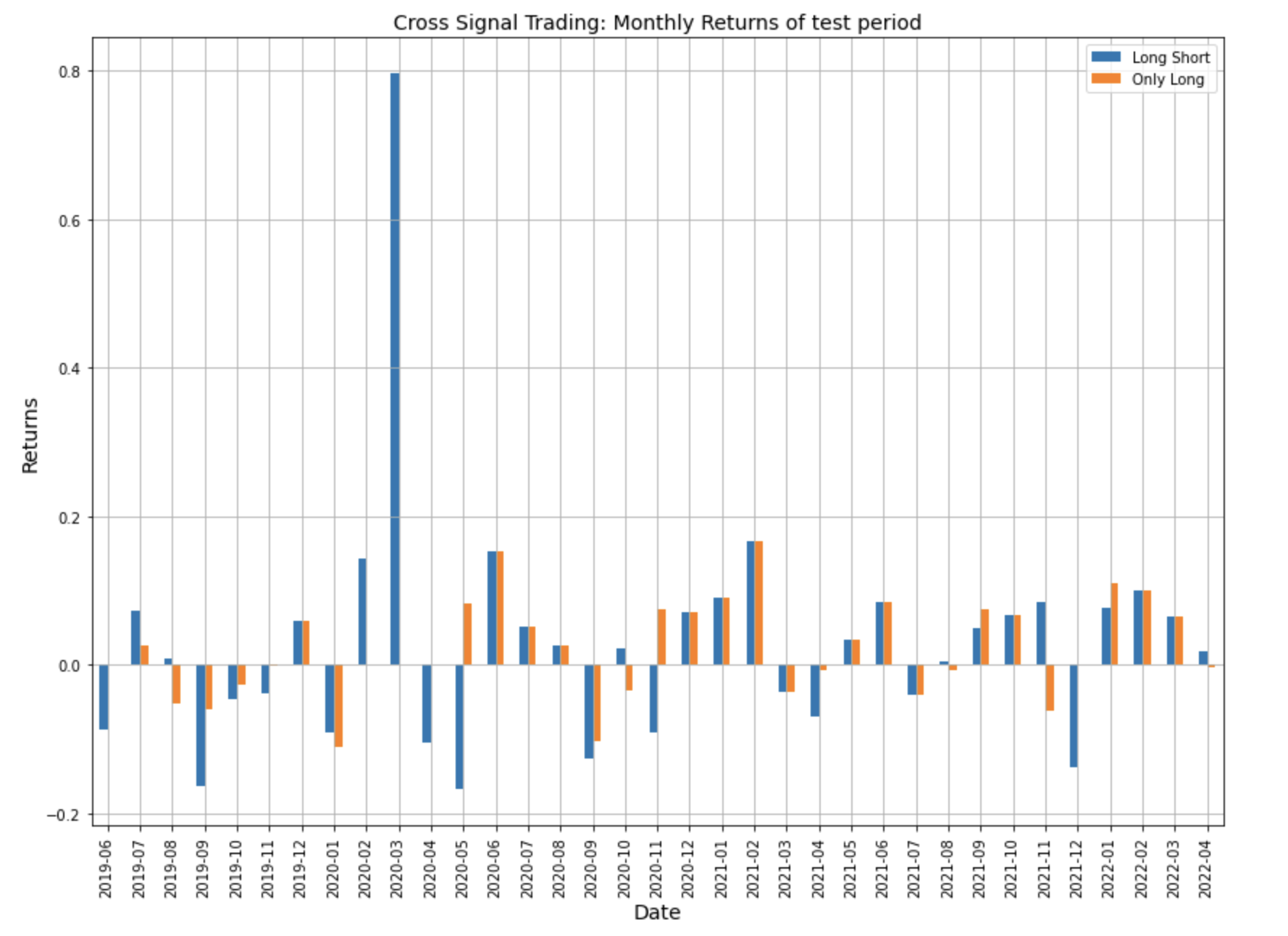}
\includegraphics[width=.45\linewidth]{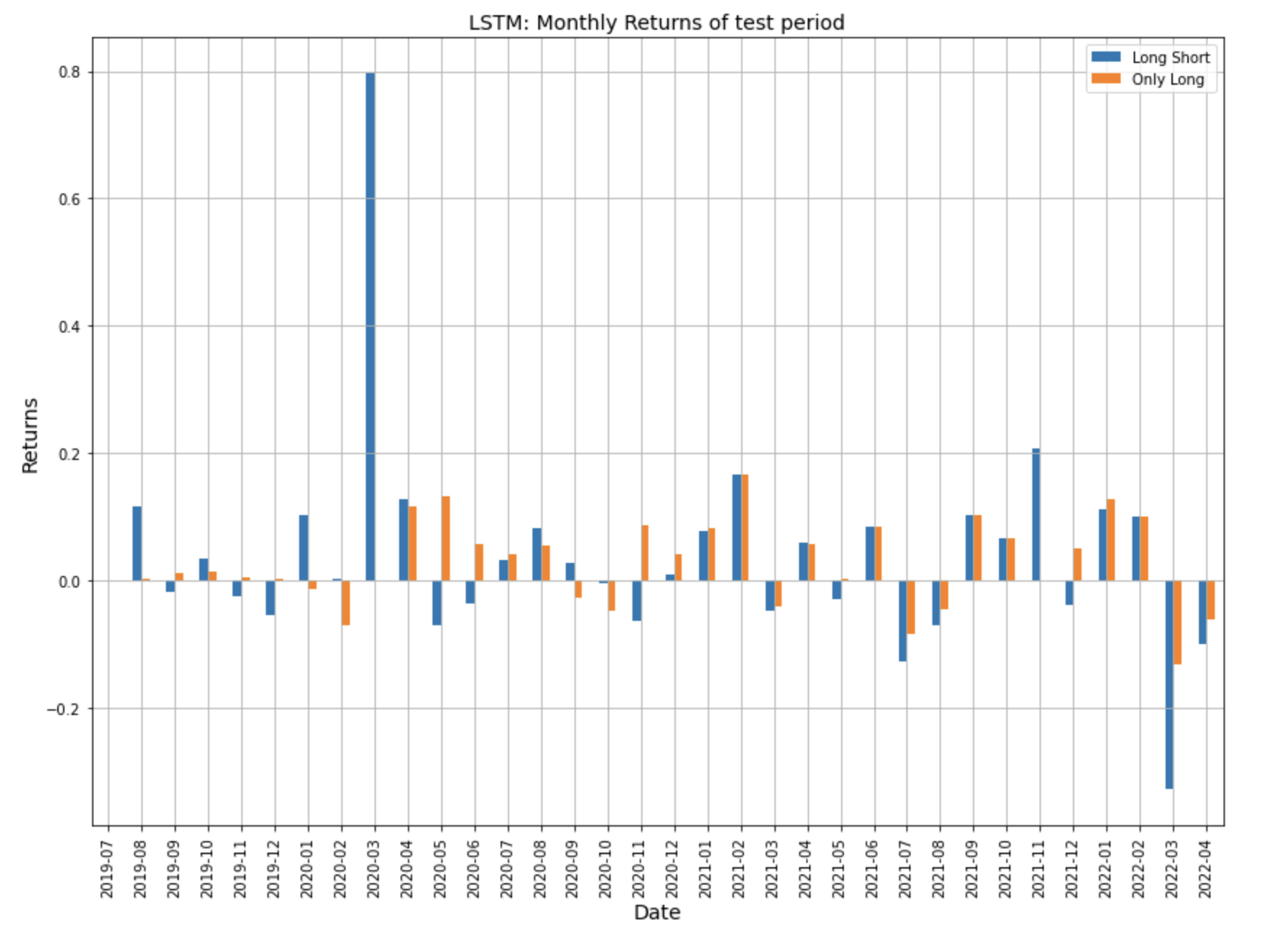}
\includegraphics[width=.45\linewidth]{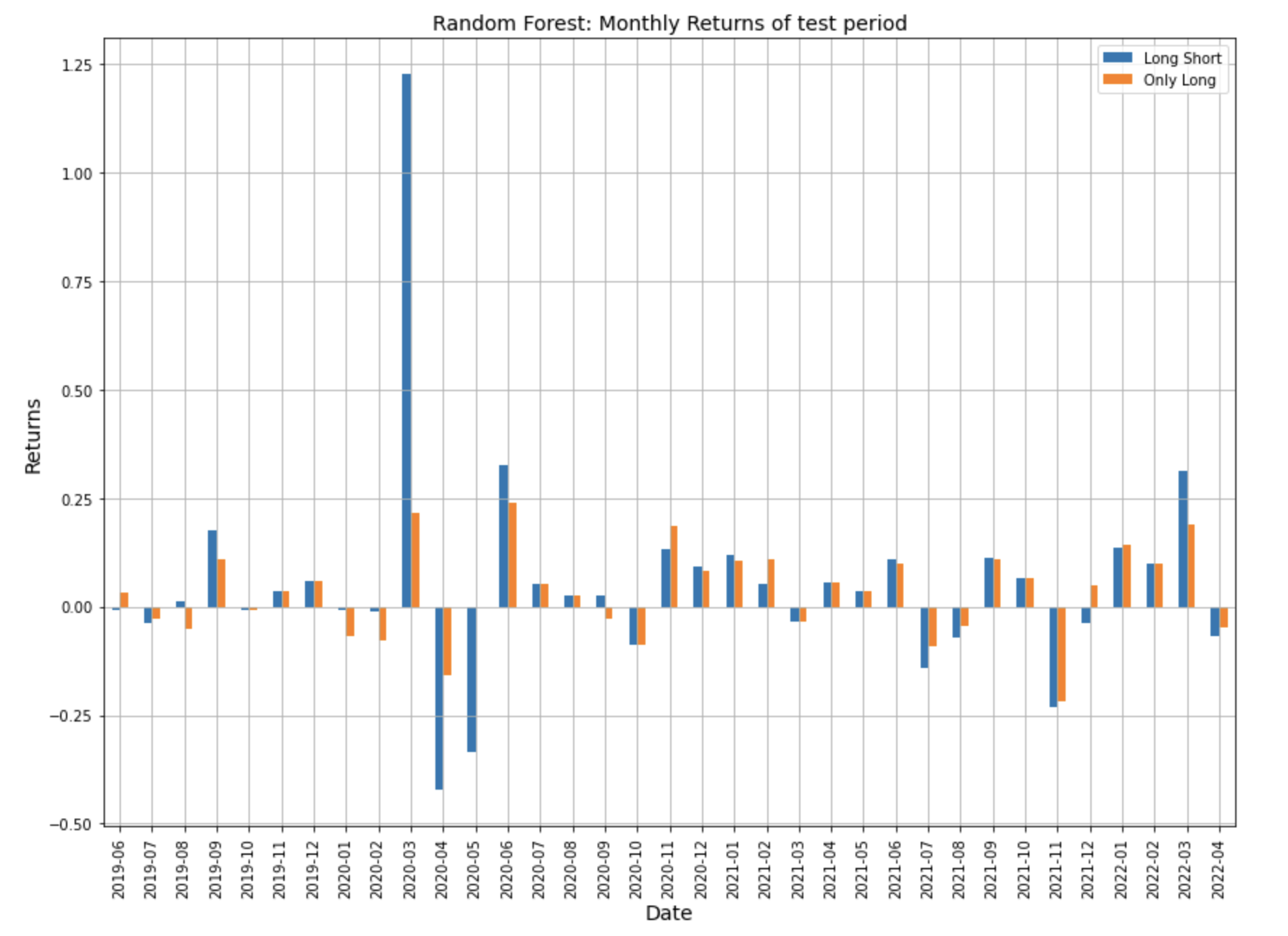}
\includegraphics[width=.45\linewidth]{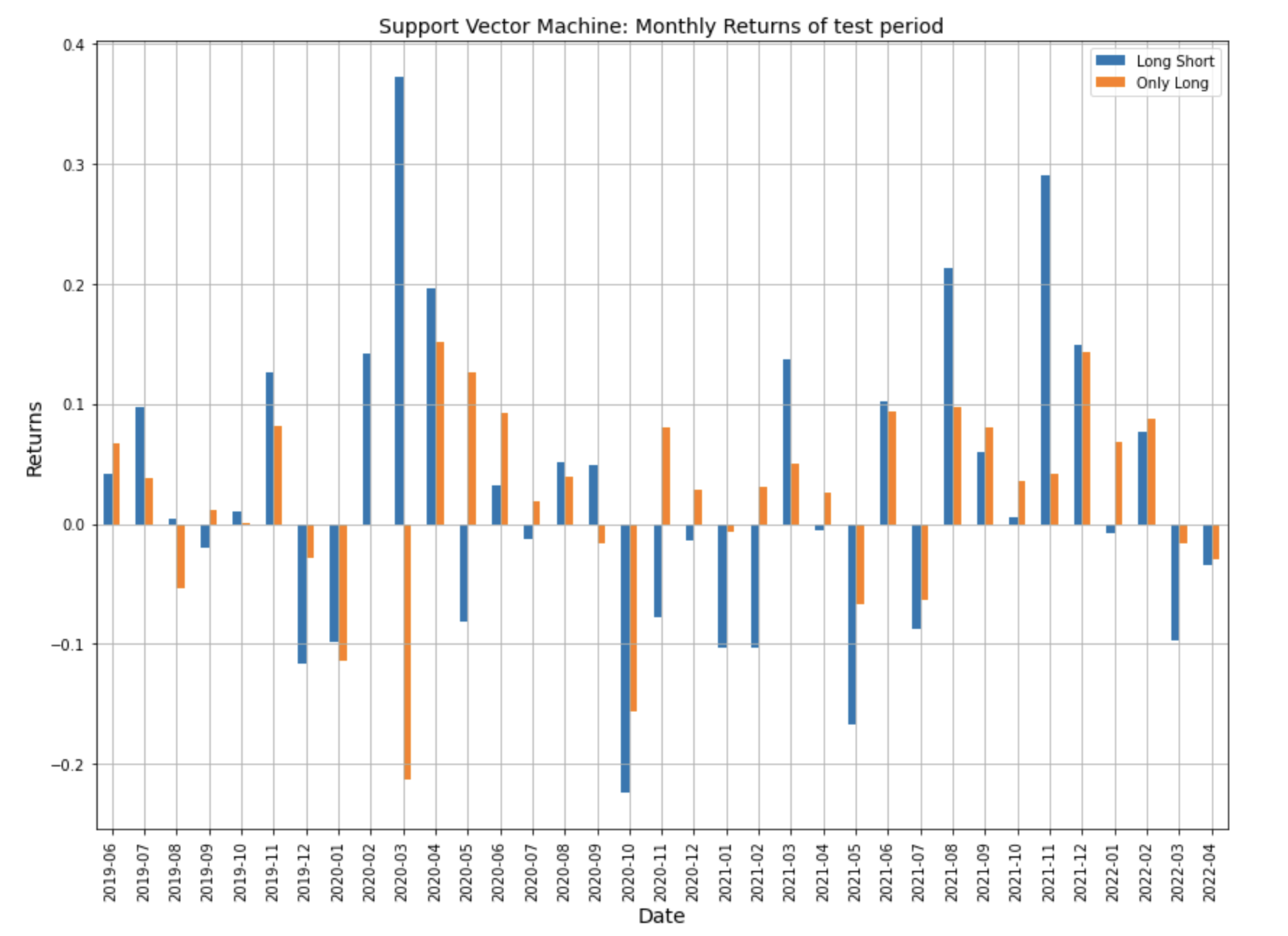}
\includegraphics[width=.45\linewidth]{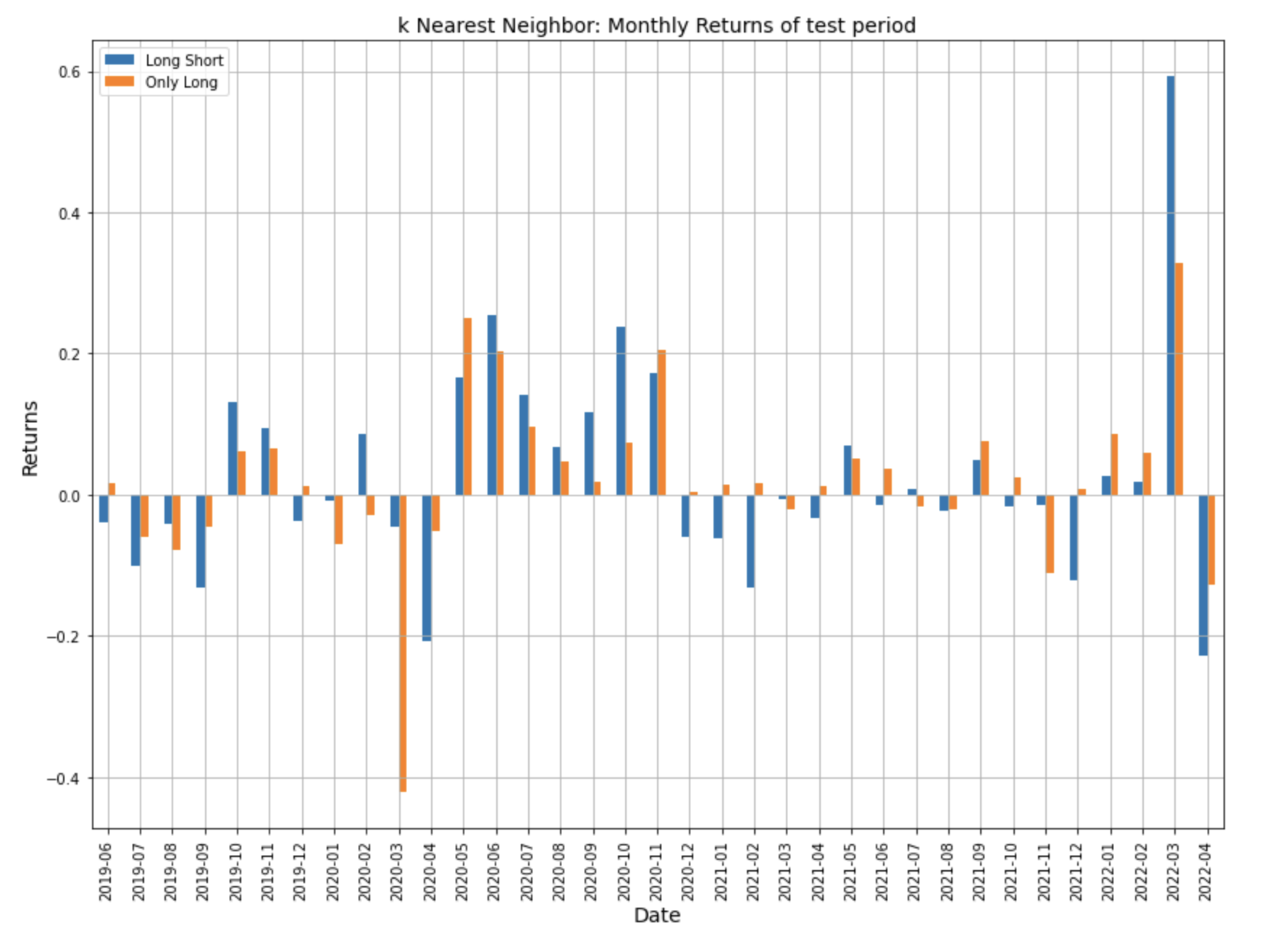}
\caption{Monthly Returns}
\label{Fig18}
\end{centering}
\end{figure}

Next, we look at the monthly returns of the individual models. Here the characteristics of the models can be analyzed. Often it is important to know in which market phases (high volatility / low volatility phases) the models perform best or worst. This can be seen in figure \ref{Fig18} . The monthly returns of the individual models reflect the global political situation on earth. In the period from 2019-12 to 2020-06 the panic on the financial markets was extremely noticeable. The large investors and private investors did not know at that time how the governments would position themselves regarding the Corona policy. The ARMA-GARCH model reflects this uncertain period very well. It can be seen that returns plummeted sharply in 2019-12. This is also seen in all other models apart from Cross Signal Trading. In this regard, the ARMA-GARCH model went downhill and then came the biggest gain. On 2020-04, the ARMA-GARCH model made almost 50\% of the total gain. The difference between the machine learning models and the conventional statistical models is that the machine learning models detected the pattern one month earlier (as early as 2020-03) than the ARMA-GARCH model. Nevertheless, the benchmark model performed very well throughout the out-of-sample period. Cross signal trading also performed very well during the Corona crisis. If we now look at the Random Forest model, we see that there were only a few losses. The model recognized very well in almost all market situations the trend patterns and also achieved the largest profit with 125\% on 2020-03. The LSTM model performed very well at the beginning of the period. It recognized the Corona pebbles very well and achieved the biggest profit of all models. However, the model loses performance in the final period. Why this happened is difficult to explain. The kNN model performed moderately. It is obvious that it did not perform well during the Corona crisis. It seems to perform better in calmer market phases than in highly volatile phases.  The SVM model, on the other hand, seems to be the most balanced model. This is because it either makes high losses or profits in every market phase. If you now look at the table \ref{Tab11} you can see the biggest gains and the maximum drawdown of the models. 

\begin{table}
\hfill
\begin{centering}
\caption{Maximum drawdown and maximum profits}
\begin{tabular}{ccccr}
\toprule[0.1pt]
         \textbf{Strategy}     & \textbf{Model}  & \textbf{Maximum Drawdown} & \textbf{Maximum Profits}\\\addlinespace
\midrule
\addlinespace
\vspace{0.1cm}
\emph{\textbf{Long-Short}}     & ARMA-GARCH      &   -0.33               &      0.96     \\
\addlinespace
\hspace{0.25cm}                & Cross Signal    &   \textbf{-0.17}              &     0.8       \\
\addlinespace
\hspace{0.25cm}                & LSTM             &   -0.33             &     0.8          \\
\addlinespace
\hspace{0.25cm}                & RF              &  -0.42     &      \textbf{1.23}      \\
\addlinespace
\hspace{0.25cm}               & SVM              &    -0.22            &   0.37  \\
\addlinespace
\hspace{0.25cm}               & kNN              &  -0.23            &  0.59     \\
\\\addlinespace
\hline
\\\addlinespace
\emph{\textbf{Only Long}}     & ARMA-GARCH      &   -0.56   & \textbf{0.53}  \\
\addlinespace
\hspace{0.25cm}               & Cross Signal    &   \textbf{-0.11}            &    0.17        \\
\addlinespace
\hspace{0.25cm}               & LSTM            &  -0.13              &     0.17         \\
\addlinespace
\hspace{0.25cm}               & RF              & -0.22              &    0.24      \\
\addlinespace
\hspace{0.25cm}               & SVM              &  -0.21                 &  0.15   \\
\addlinespace
\hspace{0.25cm}               & kNN             &  -0.42             &  0.33    \\
\addlinespace

\bottomrule[0.1pt]\addlinespace[2pt]
\label{Tab11}
\end{tabular}\par
\end{centering}
\end{table}

If we now look at the largest losses and the largest profits, we see that the RF model has made the largest profit and the largest loss in the long-short strategy. Despite the loss, the RF model has the highest profit factor overall. In the Only Long strategy, the ARMA-GARCH model made the highest monthly profit. However, the model also suffered the largest loss. The risk can be considered in different ways, if we look at the model with the lowest loss, we see that it is the Cross Signal strategy, in the Long-Short strategy it has reached a maximum loss of -0.17 and in the Only Long strategy it has reached a maximum loss of -0.11.

\newpage

\newpage

\subsubsection{Extreme value analysis}\label{Extrem_Val}

\begin{figure}[ht]
\begin{centering}
\includegraphics[width=.7\linewidth]{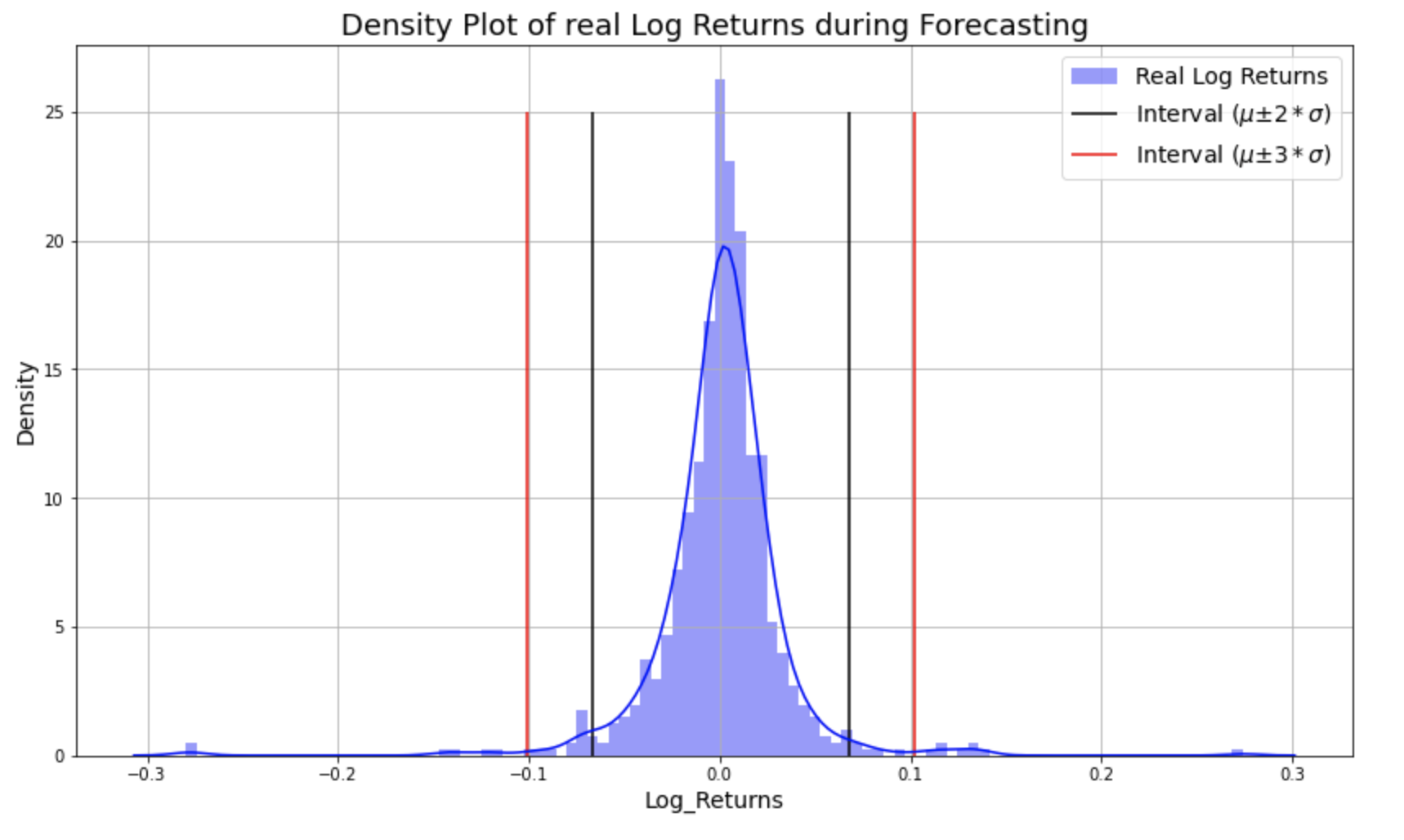}
\caption{Log Return of Prediction Horizont}
\label{Fig19}
\end{centering}
\end{figure}

In order to explain the performance of the individual models, one has to take a closer look at the accuracy of the extreme values. To see if a specific model performs better in the extreme values or, in other words, makes better decisions. In figure \ref{Fig19} the log returns of the prediction horizon are shown. In addition two intervals are visible. They can be approximated as (in black) the 95\% and (in red) the 99\% quantile. To analyze how the models perform in these quantiles, the values outside these quantiles have to be filtered out and then the accuracy has to be determined with the filtered values. 

In the table \ref{Tab12} the respective accuracy's of the models and their quantiles are visible. With this analysis it is now possible to explain why the Random Forest model performs best. It can be seen that the Random Forest model performs very well in the extreme values. This means that if a good return value is seen the next day, the model is more likely to make a correct prediction. The model allows better entries with large returns, so days with extreme return values are very profitable. The LSTM and ARCH GARCH models also perform very well in these quantiles. This is also evident in the performance. Models that do not perform well at these extreme values are therefore also less profitable. 

\begin{table}
\hfill
\begin{centering}
\caption{Extreme values accuracy} 
\begin{tabular}{ccccr}
\toprule[0.1pt]
         \textbf{Model}  & \textbf{Accuracy regular values} & \textbf{Accuracy extreme values (5\%)} & \textbf{Accuracy extreme values (1\%)} \\\addlinespace
\midrule
 ARMA-GARCH &   51.3\% &  52.78\%  &  \textbf{66.67\%}  \\

\addlinespace
\hspace{0.25cm} LSTM & 51.52\%  & 54.84\%  & 64.29\%        \\
\addlinespace
\hspace{0.25cm} RF  & \textbf{54.26\%} &   \textbf{61.11\%}  &  60.0\% \\
\addlinespace
\hspace{0.25cm} SVM & 48.91\% & 47.22\%  & 46.67\% \\
\addlinespace
\hspace{0.25cm} kNN & 52.53\% & 58.33\%   & 46.67\% \\
\addlinespace

\bottomrule[0.1pt]\addlinespace[2pt]
\label{Tab12}
\end{tabular}\par
\end{centering}
\end{table}

\newpage

\subsubsection{Cross Validation}

\begin{figure}[ht]
\begin{centering}
\includegraphics[width=.7\linewidth]{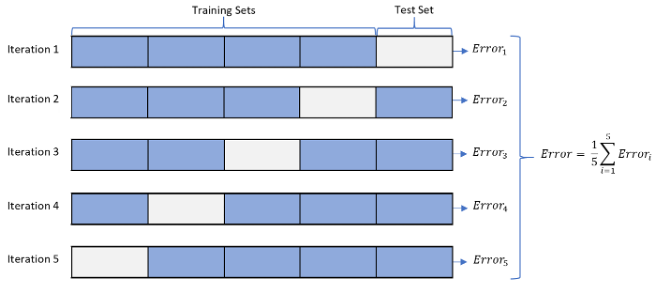}
\caption{k-Fold Cross Validation \cite{patro_2021}}
\label{Fig20}
\end{centering}
\end{figure}

Cross validation is used to obtain meaningful misclassification rate. The procedure is the same as for the original model. One determines a training and a test data set, trains the model with the training data and then makes a prediction using the test data set. For cross validation, the data is cut into k pieces of equal size. In each run, one of these pieces is defined as the test data set and the remaining data as the training data set. Then the model is trained k times and tested using test data. Thus, k predictions are obtained, which can then be quantified using a measure such as (MSE, Accuracy, MAPE, etc.). Finally, all these values can be averaged out to get a better measure and to avoid that the choice of data for test and training data was not randomly good or bad. In figure \ref{Fig20} is an illustration of the k-Fold Cross Validation. \cite{stern_2018}

In this work, the k-fold cross validation was performed with k $= 5$. It should be noted that the data is time dependent, since they are time series data. The cross validation can be performed if the k-folds keep their order. With the cross validation the robustness of the model can be tested. Nevertheless, the practical relevance is not given because with time series one wants to predict the future data by means of past data. For all models, the Sharpe ratios of the two strategies were calculated. This was also done for the CV, in addition the accuracy of the predictions for each run was determined. 

\newpage

\begin{table}
\hfill
\begin{centering}
\caption{k-Fold Cross Validation values}
\begin{tabular}{cccccccr}
\toprule[0.1pt]
               &     \textbf{k = 1} & \textbf{k = 2} & \textbf{k = 3} &  \textbf{k = 4} & \textbf{k = 5} & \textbf{Mean}\\\addlinespace
\midrule
\emph{\textbf{RF}}&         &         &         &         &         \\
\addlinespace
\hspace{0.25cm} Sharpe Ratio Only Long & 1.18 & 0.4 & -0.18 & 0.21 & 1.18 & \textbf{0.57}\\
\addlinespace
\hspace{0.25cm} Sharpe Ratio Long Short &  1.7  &  0.27 & 0.5  &  0.06 & 1.15 & \textbf{0.74} \\
\addlinespace
\hspace{0.25cm} Accuracy in \% & 53.22 & 52.13 & 49.66 & 48.7 & 54.73 & \textbf{51.69}\\
\addlinespace
\vspace{0.1em} \\ \emph{\textbf{kNN}}&         &         &         &                 \\
\addlinespace
\hspace{0.25cm} Sharpe Ratio Only Long & 1.59 & 0.65 & -0.34 & 0.85 & 0.57 & \textbf{0.66} \\ 
\addlinespace
\hspace{0.25cm} Sharpe Ratio Long Short & 2.01 & 0.51 & 0.33 & 0.99 & 0.6 & \textbf{0.89} \\
\addlinespace
\hspace{0.25cm} Accuracy in \% & 53.36 & 52.54 & 50.75 & 51.99 & 52.95 & \textbf{52.32} \\
\addlinespace \\ \emph{\textbf{SVM}}&         &         &         &                 \\
\addlinespace
\hspace{0.25cm} Sharpe Ratio Only Long & 0.06 & 1.02 & -0.58 & 0.4 & 0.78 & \textbf{0.33} \\
\addlinespace
\hspace{0.25cm} Sharpe Ratio Long Short & -0.02 & 0.87 & -0.07 & 0.21 & 0.58 & \textbf{0.32} \\
\addlinespace
\hspace{0.25cm} Accuracy in \% & 51.17 & 51.31 & 52.96 & 50.81 & 49.24 & \textbf{51.11} \\
\addlinespace \\ \emph{\textbf{LSTM}}&         &         &         &                 \\
\addlinespace
\hspace{0.25cm} Sharpe Ratio Only Long & -0.19 & 0.72 & 0.41 & 1.55 & 1.03 & \textbf{0.7}\\ 
\addlinespace
\hspace{0.25cm} Sharpe Ratio Long Short & -0.61 & 0.76 & 0.17 & 1.86 & 0.96 & \textbf{0.63} \\
\addlinespace
\hspace{0.25cm} Accuracy in \% & 50.49 & 48.68 & 50.28 & 47.71 & 45.91 & \textbf{48.61} \\
\addlinespace

\bottomrule[0.1pt]\addlinespace[2pt]
\label{Tab13}
\end{tabular}\par
\end{centering}
\end{table}

In the table \ref{Tab13} all results of the CV are shown including the arithmetic mean. Comparing the different models with each other, it is obvious that the classification models perform better than the regression models. The k-NN model with the long short strategy has the best mean Sharpe Ratio and this model also has the best mean Accuracy. You can see that the Accuracy is robust for all models, but the Sharpe Ratios are very different. Many Sharpe Ratios are negative, which indicates a mean loss. If we look at the Sharpe Ratios of k = 5, we see that they all have positive values. The k-fold k $= 5$ is also the regular way of working with time series. The model gets past data to train to predict future data. 

\newpage

\subsection{Analysis of results}

The Random Forest model shows better results than other models as it is able to correctly predict extreme returns as well as smaller returns. ARMA-GARCH and LSTM also showed very good results. These three models are able to capture the long-term dependencies in the time series and are more suitable for predicting Brent Crude Oil Future. The table \ref{Tab14} shows the Experimental Sharpe Ratios and the Profit Factors. This table confirms that machine learning models are able to perform very well on time series. 

\begin{table}
\hfill
\begin{centering}
\caption{Performance analysis}
\begin{tabular}{ccccr}
\toprule[0.1pt]
         \textbf{Strategy}     & \textbf{Model}  & \textbf{Sharpe Ratio} & \textbf{Profit Factor}\\\addlinespace
\midrule
\vspace{0.1cm}
\emph{\textbf{Buy and Hold}}   &        -         &   0.33                &      1.62     \\
\hline
\addlinespace
\vspace{0.1cm}
\emph{\textbf{Long-Short}}     & ARMA-GARCH      &   0.88               &      3.82     \\
\addlinespace
\hspace{0.25cm}                & Cross Signal    &  0.69               &   2.83         \\
\addlinespace
\hspace{0.25cm}                & LSTM             & 0.89               & 3.67              \\
\addlinespace
\hspace{0.25cm}                & RF              & \textbf{1.15}       &   \textbf{5.77}           \\
\addlinespace
\hspace{0.25cm}               & SVM              & 0.59               & 2.47    \\
\addlinespace
\hspace{0.25cm}               & kNN              & 0.60              & 2.48      \\
\\\addlinespace
\hline
\\\addlinespace
\emph{\textbf{Only Long}}     & ARMA-GARCH      &   0.77               &      2.55      \\
\addlinespace
\hspace{0.25cm}                & Cross Signal    &  0.95                &   2.21         \\
\addlinespace
\hspace{0.25cm}                & LSTM             & 0.95               & 2.43              \\
\addlinespace
\hspace{0.25cm}                & RF              & \textbf{1.18}       &   \textbf{3.13}      \\
\addlinespace
\hspace{0.25cm}               & SVM              & 0.76               & 2.07    \\
\addlinespace
\hspace{0.25cm}               & kNN              & 0.57            & 2.05     \\
\addlinespace

\bottomrule[0.1pt]\addlinespace[2pt]
\label{Tab14}
\end{tabular}\par
\end{centering}
\end{table}

%% file: 5_Conclusion.tex
\section{Conclusion and outlook}\label{Section: Conclusion}

This thesis explored how ML methods compare to already successfully established standard methods such as cross signal trading and ARMA-GARCH models in the crude oil market. First, in our literature review, we found that different machine learning methods have their strengths in different areas. Based on these findings, we evaluated the most popular methods in time series forecasting and applied them to Brent crude oil futures. These were compared with the cross signal trading strategy and the ARMA-GARCH model. While the properties of ARMA-GARCH models are suitable for modeling a variable variance, the ML methods can be beneficial for pattern recognition in the data. To this end, measurements were made on the quality of the predictions of each model. The Random Forest was characterized by a very high true-positive rate (recall), which set it apart from the other models. Likewise, the final performance result provided conclusions about the robustness of the individual models. It could be shown in which market phases the models made strong or poor trading decisions. In the case of the kNN model, which failed to detect the Covid-19 crisis, it was shown that not every ML model responded well to different market phases. Comparing this with the RF model, it can be seen that the most significant monthly gain was achieved there across all models.
The conventional statistical model - ARMA-GARCH - performed surprisingly well and was on par with the ML models. Nevertheless, the extreme values were analyzed in more detail. It was found that the RF model made the best decisions at the 95\% interval level. This shows that the RF model continuously made better decisions than all other models during a volatile market period, which is reflected in the total return. Surprisingly, the ARMA-GARCH model made the best decisions at the 99\% level but performed worse on average during periods when the market was "calmer." One direction for future work will be the volatility of stock time series. One difficulty in predicting stock markets arises from their non-stationary behavior. It would be interesting to see how machine learning models perform on denoised data or how these models perform in other crude oil markets.